\documentclass[twocolumn,secnumarabic,amssymb, nobibnotes, aps, pre, floatfix, nofootinbib]{revtex4-1}

\usepackage[utf8]{inputenc}
\usepackage[english]{babel}
\usepackage[colorlinks=true,allcolors=blue]{hyperref}%
\usepackage{graphicx}
\usepackage[dvipsnames]{xcolor}
\usepackage{epsfig}
\usepackage{amsmath}
\usepackage{mathtools}
\usepackage{booktabs}
\usepackage{float}
\usepackage{lipsum}

\newcommand{\dert}[2]{ {{\rm d}#1\over {\rm d}#2}}
\begin{document}

\title{Diffusion and pattern formation in spatial games}%

\begin{abstract}
    Diffusion plays an important role in a wide variety of phenomena, from bacterial quorum sensing to the dynamics of traffic flow. While it generally tends to level out gradients and inhomogeneities, diffusion has nonetheless been shown to promote pattern formation in certain classes of systems. Formation of stable structures often serves as a key factor in promoting the emergence and persistence of cooperative behavior in otherwise competitive environments, however an in-depth analysis on the impact of diffusion on such systems is lacking. We therefore investigate the effects of diffusion on cooperative behavior using a cellular automaton (CA) model of the noisy spatial iterated prisoner’s dilemma (IPD), physical extension and stochasticity being unavoidable characteristics of several natural phenomena. We further derive a mean-field (MF) model that captures the 3-species predation dynamics from the CA model and highlight how pattern formation arises in this new model, then characterize how including diffusion by interchange similarly enables the emergence of large scale structures in the CA model as well. We investigate how these emerging patterns favors cooperative behavior for parameter space regions where IPD error rates classically forbid such dynamics. We thus demonstrate how the coupling of diffusion with non-linear dynamics can, counter-intuitively, promote large scale structure formation and in return establish new grounds for cooperation to take hold in stochastic spatial systems.
\end{abstract}

\author{Alexandre Champagne-Ruel}%
\affiliation{Département de physique, Université de Montréal, Canada}
\email[Corresponding author: ]{alexandre.champagne-ruel@umontreal.ca}

\author{Sascha Zakaib-Bernier}
\affiliation{Département de physique, Université de Montréal, Canada}

\author{Paul Charbonneau}
\affiliation{Département de physique, Université de Montréal, Canada}

\date{July 1, 2024}%
\maketitle

\section{Introduction}
\label{sec:intro}

Cooperative phenomena are ubiquitous in nature. While biological evolution hinges on the principle of selection, which also implies competition, instances of cooperation are widespread both at small~\cite{turner_prisoners_1999, bohl_evolutionary_2014, hong_competition_1992, sievers_self-replication_1994, lee_autocatalytic_1997} and larger~\cite{dugatkin_cooperation_1997, dugatkin_guppies_1988, heinsohn_complex_1995, kay_evolution_2020, lombardo_mutual_1985, milinski_tit_1987} scales. The evolution of mitochondrial cells is often cited as one prime example of mutually beneficial behavior~\cite{sagan_origin_1967,margulis_symbiosis_1981}, and cooperation is believed to have been instrumental both in the emergence of life on Earth~\cite{kauffman_autocatalytic_1986, hordijk_detecting_2004, hordijk_history_2019} and, more generally, in major evolutionary transitions throughout history that would have necessitated the establishment of new forms of cooperation \cite{queller_cooperators_1997, szathmary_toward_2015, massey_origin_2018}.

Game theory is the framework of choice for examining the emergence of cooperation in nature, a problem that has motivated extensive analysis across various fields~\cite{colman_game_2013, myerson_game_2013, watson_strategy_2016}, from ecology~\cite{sigmund_games_1995} to computational neuroscience~\cite{keinan_axiomatic_2006}---and more recently in the fields of physics and biochemistry~\cite{schuster_use_2008}. One way to study cooperative phenomena is through the formalism of the Prisoner’s Dilemma (PD)~\cite{fudenberg_game_1991}: in this game, two players must decide whether to cooperate or defect, without prior knowledge of their opponent's choice. The PD’s formalism can be defined as a score matrix that highlights the tension between rational, selfish behaviors and mutual cooperation that benefits the group as a whole (Table~\ref{tab:pd}). A game is classified as PD when the Reward $R$ (both players cooperate), the Punishment $P$ (they both defect), the Temptation $T$ (one exploits its opponent) and the Sucker’s payoff $S$ (one is being exploited) satisfy the constraint $T>R>P>S$~\cite{nowak_evolutionary_2006}. 

Numerous competitive scenarios align with the structure of the PD. This includes interactions among viruses~\cite{turner_prisoners_1999} or fish~\cite{dugatkin_cooperation_1997} to the dynamics of WWI trench warfare~\cite{trivers_social_1985}. As such, the PD stands as the model of choice in studying what factors drive, influence, and favor the emergence of cooperative behavior.

\begin{table}
\begin{ruledtabular}
\begin{tabular}{ccc}
           & Cooperation & Defection \\ \hline
Cooperation & 3 ($R$)         & 0 ($S$)         \\
Defection & 5 ($T$)         & 1 ($P$)         \\
\end{tabular}
\end{ruledtabular}
\caption{\label{tab:pd}Canonical score matrix defining the Prisoner’s Dilemma (PD). Two players, without knowing their opponent’s choice in advance, can either choose to cooperate or defect. While exploitation (i.e., defecting while the opponent cooperates) leads to the highest score for one player, the \textit{average} score between both players is higher when they simultaneously cooperate. The PD thus highlights the tension between rational, selfish behavior, and an alternative that benefits the group as a whole.}
\end{table}

In real world scenarios, players rarely encounter one another only once. Accordingly, the PD can be extended to account for \textit{repeated} interaction between players. The Iterated Prisoner’s Dilemma (IPD) defines a game of length $M$ wherein two players will repeatedly choose between cooperation and defection, guided by a specific strategy. Strategies range from straightforward ones such as "Always Cooperate" (ALLC) and "Always Defect" (ALLD) to reactive ones such as "Tit-for-Tat" (TFT), which consistently mirrors the opponent’s last move. Other well-known strategies include "Playing at Random", "Win-Stay-Lose-Shift", "Generous Tit-for-Tat", among others~\cite{delahaye_complex_1995, nowak_evolutionary_2006, jurisic_review_2012}. 

Despite the abundant literature aiming to identify the optimal strategy for the IPD~\cite{haider_using_2005, sandholm_multiagent_1996}, TFT consistently emerges as a successful contender against a wide array of opponents. The simple act of mirroring the opponent’s last move remains unparalleled in versatility, even when strategy makers know in advance that they would be playing against it~\cite{axelrod_evolution_1981, axelrod_evolution_1984}. Just as the PD is considered the cornerstone for studying cooperative phenomena, TFT is regarded as its pivot~\cite{nowak_tit_1992}. Therefore, we explore in what follows the dynamics between the most basic IPD strategies---ALLC, ALLD and TFT---as they are more likely to emerge spontaneously in such a competitive setting.

\begin{figure}
    \includegraphics[width=\linewidth]{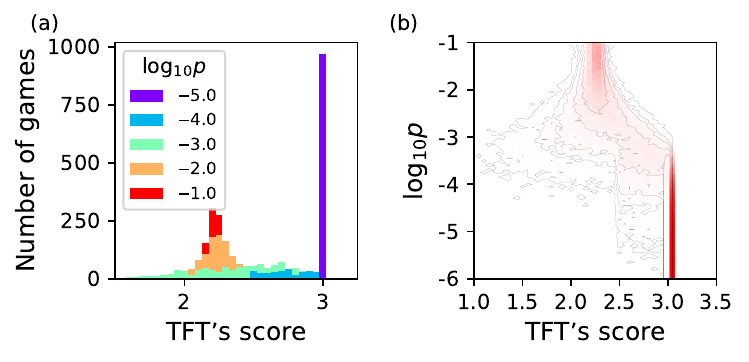}
    \caption{Mean score for two TFT players competing against each other for $M=1000$ moves, averaged over a statistical sample of $1000$ IPD simulations. (\textbf{a}) As the error rate increases from a small value ($\log_{10} p=-5$, purple bars) towards its maximum ($\log_{10} p=-1$, red bars), the score distribution shifts from a narrow distribution with mean $\mu\sim3.0$ to a broader distribution where $\mu\sim2.25$, reflecting the average score across all four move combinations. (\textbf{b}) Two-dimensional histogram showing the distribution of error rates for the same TFT players, with the vertical axis representing variations in the error rate. The mean score peaks when $p$ is minimal (lower region) and then shifts towards 2.25 (upper region) as the error rate approaches the critical value $p_{\rm c}\sim M^{-1}=10^{-3}$, at which point players commit at least one error per game.}
    \label{fig:hists}
\end{figure}

In the context of the IPD, one important limitation of TFT is however its susceptibility to the presence of \textit{errors}. Given that statistical physics asserts all physical systems experience fluctuations at non-zero temperatures~\cite{reif_fundamentals_2009}, this parameter can become crucial across a wide variety of systems. If one of two cooperating TFT players mistakenly defects, they will both lock onto a sequence of mutual defection. As illustrated on Fig.~\ref{fig:hists}, the average score of TFT playing the IPD against itself will quickly plummet with an increase in the error rate~\cite{nowak_evolutionary_2006}. Additionally, the absence of an error-correction mechanism in TFT makes it susceptible to being superseded by \textit{unconditional} cooperators such as ALLC. Unlike TFT, ALLC doesn’t engage in similar mutual defection cycles at non-zero error rates and can dominate once their predators are removed.

Extensive research has revealed how several mechanisms can influence the emergence of cooperation in challenging environments, such as payoffs~\cite{szabo_evolutionary_1998, lim_scale-invariant_2002, szabo_phase_2005}, noise~\cite{chen_interaction_2008, szolnoki_selection_2009, fant_stable_2023}, or topology~\cite{szabo_spatial_2000, vukov_cooperation_2006, amaral_stochastic_2016}. Likewise, the role of mobility as a phenotypical characteristic (i.e., under direct control of the agents or their strategy) has been subjected to comprehensive analyses~\cite{dugatkin_cooperation_1997, enquist_evolution_1993, ferriere_invading_1995, aktipis_know_2004, hamilton_contingent_2005, le_galliard_adaptive_2005, helbing_outbreak_2009, meloni_effects_2009, smaldino_movement_2012, yang_random_2023}. However, one related but prevalent feature spanning a wide variety of physical systems, molecular diffusion---i.e., the Brownian-like (random) movement that characterizes the particles of any physical system at non-zero temperature---, has been comparatively less explored in this context. Systems that both rely on cooperative behaviors during their evolution and simultaneously experience the physical effects of diffusion are common in nature and across various scientific domains. Significant examples include many biological systems, such as that of bacterial colonies engaging in quorum sensing~\cite{ben-jacob_cooperative_2000} and where diffusion influences the dynamics of signaling~\cite{bryant_physical_2023}, of ant populations that act in coordinated actions and also rely on the diffusion of pheromones in the environment to do so~\cite{holldobler_ants_1990}, or of bird flocking during the formation of large patterns~\cite{cattivelli_modeling_2011}. Many other instances highlight the pivotal influence of diffusion, from cooperative chemical species in catalytic reaction systems~\cite{turing_chemical_1952} to the coordinated action of vehicles~\cite{nagel_emergent_1995} or pedestrians~\cite{von_schantz_spatial_2015, yanagisawa_coordination_2016} in traffic flow that manifest diffusion-like dynamics. Previous analyses were performed that included molecular diffusion in the context of simulations of the IPD~\cite{ferriere_evolution_1996, vainstein_does_2007, sicardi_random_2009, droz_motion_2009, cheng_motion_2010, suzuki_oscillatory_2011, yang_universal_2011, gelimson_mobility_2013, vainstein_percolation_2014}, however an extensive investigation of how pattern formation characterizing diffusion-like processes affect the dynamics of the spatial IPD is still lacking.

In what follows, we aim to address this question about the influence of diffusion on cooperation by characterizing its effect on the interplay between cooperators and defectors. Using both a deterministic cellular automaton (CA) model (Section~\ref{sec:IPD}) and a mean-field (MF) model (Section~\ref{sec:mean-field-model}) of the spatial iterated prisoner’s dilemma, we investigate how pattern formation affects the dominance of strategies within the IPD. While diffusion usually tends to level out gradients and inhomogeneities, it has long been known that coupling it to nonlinear dynamics can give rise to pattern formation~\cite{turing_chemical_1952}: our results indicate that this phenomena further promotes cooperation (Section~\ref{sec:diffusion-discrete}), which could thus explain how cooperative behavior arises in otherwise adverse environments where diffusion takes place.

\section{Cellular automaton model for a 3-strategy spatial game}
\label{sec:IPD}

Spatial constraints like topology or dimensionality exert significant influence on the dynamics of evolutionary games~\cite{fu_invasion_2010, lindgren_evolutionary_1994, nowak_games_1999}. Given that these are inherent in various biological, ecological, and social phenomena, we investigate the iterated prisoner’s dilemma using numerical simulations on a 2D lattice of sufficient size ($128\times128$) to reduce finite-size effects such as stochastic extinction of strategies.

\begin{figure}
    \includegraphics[width=\linewidth]{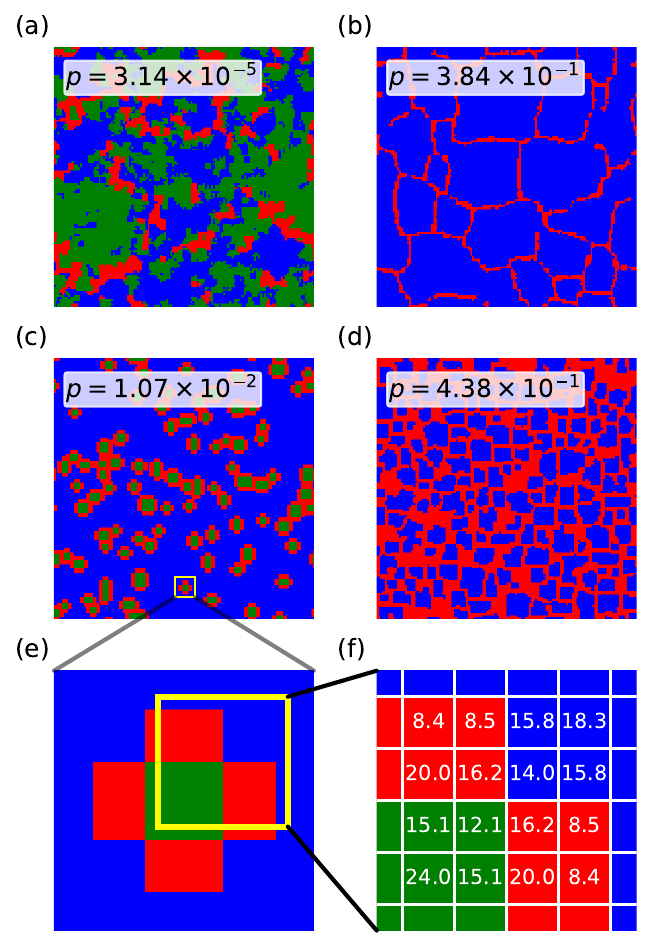}
    \label{fig:lipd-snaps}
    \caption{Final lattice state for cellular automaton simulations with varying error rates $p$. See also Appendix~\ref{sec:anims} for animated versions of Panels (a) and (c). Simulations were carried out on a Cartesian lattice of size $128\times128$, with periodic boundaries and random initial conditions, over $T=500$ model iterations. Games of the iterated prisoner’s dilemma of $M=2000$ moves were played. \textbf{(a)} Fractal patterns are formed, across which strategies propagate through Fisher waves \protect\citep{fisher_wave_1937}. \textbf{(b)} Filament-like structures of defectors (ALLD, red) surviving among a TFT-filled lattice (blue). \textbf{(c)} Clusters of cooperative strategies (ALLC, green) surrounded by parasites (ALLD, red), in a TFT-filled lattice (blue). \textbf{(d)} Mottled patterns interweaving cooperators (TFT, blue) and defectors (ALLD, red). \textbf{(e-f)} Close-up view of stable ALLC clusters surrounded by ALLD strategies. The central strategies (ALLC) survive through mutual cooperation, yielding the highest scores in their Moore neighborhood, while the parasites can persist by defecting against ALLC. The scores shown on panel (f) are averages taken over a statistical ensemble of $10^4$ games.}
\end{figure}

In the deterministic cellular automaton (CA) model, simulations proceed by distributing equal populations of each strategy randomly across the lattice. At each iteration $t_i$ in the model, every player on the grid engages in $M=2000$ rounds of the prisoner’s dilemma with the eight players in its Moore neighborhood. After these rounds, the players will adopt the highest-scoring player’s strategy in this neighborhood if it outperforms theirs. These replacements are deterministic and sequential, but they proceed in a random order to eliminate any spatial bias. The simulation runs for $T=500$ iterations, ensuring a relaxation of the system’s dependence on the initial conditions for every realization of the model.

The repeated interaction between players, whereby they engage $M$ times in the PD, is a key property of the model. These repeated interactions are a reflection of the fact that systems relying both on cooperative behaviour and experiencing diffusion are not well-mixed. Hence, in these systems the characteristic diffusion time-scale is typically longer than the time of interaction between the individuals, which increases the likelihood of repeated encounters. While parasitic strategies like ALLD will inevitably dominate cooperators like ALLC in single interactions, repeated encounters enable reactive strategies such as TFT to reciprocate any potential cooperative behaviour. Furthermore, selecting a larger value for $M$ allows for the exploration of a broader range of error rates while maintaining the statistical significance of the players’ mistakes over the total number of games.

Integrating game playing errors in the CA model, which reflects the inevitable stochasticity of natural processes, further expands the variety of behaviors the simulation displays. Specifically, in each of the $M$ rounds of the IPD, players have a probability $p$, drawn from a uniform distribution, of taking the opposite action (i.e., defecting when supposed to cooperate, or vice-versa). The spatial IPD has indeed been demonstrated to produce diverse spatial patterns the error rate is varied: even for low values of $p$ ($-5\lesssim \log_{10} p \lesssim -3$), novel behavior emerges, as depicted on Fig.~\ref{fig:lipd-snaps}a: ALLC gains the ability to evade ALLD by propagating into areas dominated by TFT. This sets off a series of wave-fronts where ALLC further replaces TFT, with a second trailing front across which ALLD replaces ALLC. Regions dominated by TFT finally expand into ALLD from behind (see Fig.~\ref{fig:lipd-snaps}a and animations in Appendix~\ref{sec:anims}). Subsequent panels on Fig.~\ref{fig:lipd-snaps} display the resulting lattices from simulations conducted with varying error rates. Formed structures thereafter include filaments of ALLD strategies interspersed within a TFT-dominated lattice, clusters of ALLC strategies encircled by ALLD or mottled patterns of defectors. These spatial constraints can foster cooperation in conditions where it would otherwise be absent~\cite{flores_cooperation_2022, nowak_spatial_1994, wang_spatial_2012}. For example, unconditional cooperators (ALLC) tend to be quickly eliminated by parasites (ALLD) unless they can form stable clusters ensuring their survival (Fig.~\ref{fig:lipd-snaps}c)~\cite{nowak_evolutionary_2006}. 

Examination of many such simulations indeed reveals that, barring very low or very high error rates, the spatiotemporal evolution of strategies can---to good approximation---be reduced to three distinct ``predatory'' interactions: (1) ALLD rapidly replacing ALLC; (2) cooperating groups of TFT gradually replacing ALLD; (3) ALLC slowly replacing TFT because of the latter's inability to correct mistakes when playing against itself at non-zero error rates. This allows the design of a mean-field model~\cite{szabo_spatial_2000} capturing this three-species predator-prey dynamics. 

\section{Mean-field model}
\label{sec:mean-field-model}

\subsection{A 3-species predator-prey model}
\label{ssec:3species}

With the three primary predation interactions identified, we can construct a model of the spatial IPD in the mean-field (MF) limit. The time evolution for the population fractions $C$, $T$, and $D$ for the three strategies ALLC, TFT and ALLD take the form:
\begin{eqnarray}
\label{eq:odeC}
\frac{{\rm d} C}{{\rm d} t}&=& \alpha CT - \gamma C D~,\\
\label{eq:odeT}
\frac{{\rm d} T}{{\rm d} t}&=&-\alpha CT + \beta T D~,\\
\label{eq:odeD}
\frac{{\rm d} D}{{\rm d} t}&=& \gamma CD - \beta T D~.
\end{eqnarray}
with $0\leq C,T,D\leq 1$. Here $\alpha$ is the predation rate of TFT by ALLC, $\beta$ the predation rate of ALLD by TFT, and $\gamma$ that of ALLC by ALLD. This triangular predatory dynamics is in fact akin to that of the Rock-Paper-Scissor game \cite{szabo_evolutionary_2007, szolnoki_cyclic_2014}.

\begin{figure}
    \includegraphics[width=\linewidth]{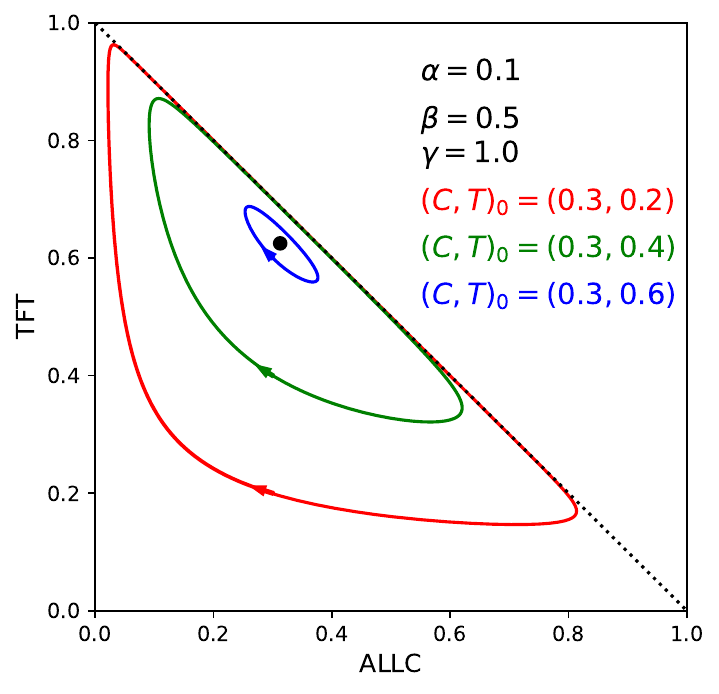}
    \caption{Representative solutions of Eqs.~(\ref{eq:ode2C})--(\ref{eq:ode2T}) in the $[C,T]$ plane, for $\alpha=0.1$, $\beta=0.5$, et $\gamma=1$ and three different initial conditions, as color-coded. The equilibrium solution is indicated by the black dot. The trajectories describe temporal oscillations of the population fractions qualitatively resembling oscillations observed in classical predator-prey dynamics or autocatalytic sets of chemical reactions, although both dynamics remain distinct (see text).}
\label{fig:1sol}
\end{figure}

Given the constraint $C+T+D=1$ one of the above ODEs is redundant and the above system can be reduced to:
\begin{eqnarray}
\label{eq:ode2C}
\frac{{\rm d} C}{{\rm d} t}&=& \alpha CT - \gamma C (1-C-T)~,\\
\label{eq:ode2T}
\frac{{\rm d} T}{{\rm d} t}&=&-\alpha CT + \beta T (1-C-T)~.
\end{eqnarray}
The equilibrium solution (${\rm d}/{\rm d}t\equiv 0$) is:
\begin{equation}
\label{eq:EqC}
C_{\rm eq}= {\beta\over s}~,\qquad
T_{\rm eq}= {\gamma\over s}~,\qquad
D_{\rm eq}= {\alpha\over s}~.
\end{equation}
with $s\equiv \alpha+\beta+\gamma$.
Figure~\ref{fig:1sol} shows three representative solutions to Eqs.~(\ref{eq:ode2C})--(\ref{eq:ode2T}), plotted as trajectories in the $[C,T]$ planes (the corresponding 3D trajectories are restricted to the wedge defined by the $C+T+D=1$ and $C=0, T=0$, and $D=0$ planes).The three solutions plotted differ only in their initial conditions, as color-coded.
These trajectories described temporal oscillations of the population fractions, qualitatively resembling oscillations observed in classical predator-prey (e.g., Lotka-Volterra) dynamics \citep{song_host-parasite_2015} or in autocatalytic (e.g., Belousov-Zhabotinsky) sets of chemical reactions \citep{kondepudi_modern_2015}. 

However, the resemblance is only superficial.
Here we have no attractors in the form of fixed points or limit cycles. Instead, for a given set of reaction rates, an infinity of stable orbits are possible, and entirely determined by the initial conditions through the invariant $C^\beta T^\gamma D^\alpha=$~constant, which can be obtained by algebraic manipulations of Eqs.~(\ref{eq:odeC})--(\ref{eq:odeD}). The distinction hinges on the fact that predator-prey and autocatalytic systems are open systems, with external reservoirs acting as sources and their magnitude determining the equilibrium solutions and character of limit cycles; while our system is closed and strictly conservative in its population size, and its equilibrium solutions are entirely determined by the predations rates.

Further insight into these differences can be obtained through 
the linearization of
Eqs.~(\ref{eq:ode2C})--(\ref{eq:ode2T}) about the equilibrium
solutions (\ref{eq:EqC}).
This yields, to first order:
\begin{eqnarray}
\label{eq:linode2C}
\frac{{\rm d} C_1}{{\rm d} t}&=&
\left[{\beta\gamma\over s}\right]C_1
+\left[{(\alpha+\gamma)\beta\over s}\right]T_1~,\\
\label{eq:linode2T}
\frac{{\rm d} T_1}{{\rm d} t}&=&
-\left[{(\alpha+\beta)\gamma\over s}\right]C_1
-\left[{\beta\gamma\over s}\right]T_1~,
\end{eqnarray}
with $C_1(t)$ and $T_1(t)$ the first order perturbations. 
These two coupled linear ODEs
can be combined in the decoupled harmonic equations:
\begin{equation}
\left(\frac{{\rm d}^2}{{\rm d} t^2}
+\frac{\alpha\beta\gamma}{s}\right)
\begin{pmatrix}
C_1 \\ T_1
\end{pmatrix}
=\begin{pmatrix}
0 \\ 0
\end{pmatrix}.
\end{equation}
These describe harmonic oscillation about the equilibrium solution,
with angular frequency $\omega=(\alpha\beta\gamma/s)^{1/2}$, and constant (small) amplitudes set by the initial condition. Our system is thus dynamically closer to an undamped nonlinear pendulum than it is to classical predator-prey systems, or to autocatalytic chemical reaction systems.

Two of the three trajectories plotted on Fig.~\ref{fig:1sol} spend a significant fraction of their time running closely along the $C+T=1$ line, implying that $D\to 0$ there. Although stochastic extinction cannot take place in this model, we do observe it in the discrete spatial IPD (CA model) described in \S\ref{sec:IPD}. If we set $D=0$, the system (\ref{eq:ode2C})--(\ref{eq:ode2T}) reduces to the single ODE:
\begin{equation}
\label{eq:odeCextD}
\dert{C}{t}= \alpha C(1-C)~,
\end{equation}
with $T=1-C$. This describes a slow drift towards the new equilibrium solution $[C_{\rm eq},T_{\rm eq}]=[1,0]$ on the timescale $\alpha^{-1}$, as observed in the IPD cellular automaton simulations when errors in game-playing occur, following stochastic extinction of ALLD.
Likewise, stochastic extinction of ALLC would lead to a drift towards $[D_{\rm eq},T_{\rm eq}]=[0,1]$ on the timescale $\beta^{-1}$.

\subsection{Estimating predation rates from CA simulation}

\begin{figure}
    \includegraphics[width=\linewidth]{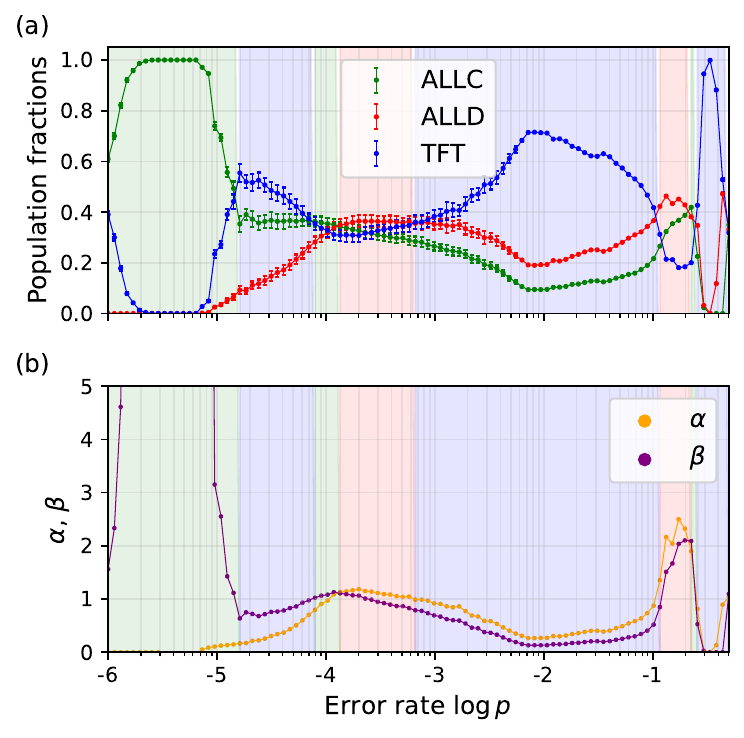}
    \caption{\textbf{(a)} Final population fractions for strategies ALLC (green), ALLD (red) and TFT (blue) playing spatial IPD games of varying error rates which were simulated using the cellular automaton model. A set of 100 values of the error rate were sampled in the interval $p\in[10^{-6},0.5]$. Error bars represent standard deviations over each statistical ensemble of 10 realizations of the model. Background hues indicate the strategy that dominates for a given error rate. \textbf{(b)} Predation rates used as parameters for the mean-field model, derived for each error rate from population fractions shown in panel (a) using Equation~\eqref{eq:alphabetaeq}.}
    \label{fig:from-p-to-alpha-beta}
\end{figure}

Using the inverse predation rate $\gamma^{-1}$ as a time unit, the equilibrium solutions (\ref{eq:EqC}) can be recast as:
\begin{equation}
\label{eq:alphabetaeq}
\alpha = {1\over T_{\rm eq}}(1-C_{\rm eq}-T_{\rm eq})~,\qquad
\beta = {C_{\rm eq}\over T_{\rm eq}}~, 
\end{equation}
allowing to estimate the dependence of predation rates on game playing error rates in the cellular automaton IPD simulations. At low error rates ($-5\leq \log_{10} p \leq -2$), this yields the hierarchy $\alpha, \beta \leq 1$. This no longer holds for regimes where the system enters absorbing states (i.e., one or more strategies goes extinct), for instance regions where error rates approach zero ($\log_{10} p\leq-5$) or where players have a high probability of making mistakes ($\log_{10} p\geq-1$). The equivalency between the error rate $p$ for the CA model and parameters $\alpha,\beta$ in the MF model is shown on Fig.~\ref{fig:from-p-to-alpha-beta}, calculated for 100 sampled values of the error rate $\log_{10} p\in[-6,-1]$ with statistical ensembles over 10 realizations of the simulation which were carried out with random initial conditions for each value of $p$.

\subsection{Spatial extension and diffusion}
\label{ssec:spatial-ext}

To investigate the effects of diffusion, we introduce two-dimensional spatiality and isotropic linear diffusion in the MF model; Equations (\ref{eq:ode2C})--(\ref{eq:ode2T}) are replaced by:
\begin{eqnarray}
\label{eq:pde2C}
\frac{\partial C}{\partial t}&=& \alpha CT - \gamma C (1-C-T)+ D_C\nabla^2 C~,\\
\label{eq:pde2T}
\frac{\partial T}{\partial t}&=&-\alpha CT + \beta T (1-C-T)+ D_T\nabla^2 T~.
\end{eqnarray}
with $C(x,y,t)$ and $T(x,y,t)$, and the diffusion coefficients $D_C,D_T$ need not be identical.

We solve Eqs.~(\ref{eq:pde2C})--(\ref{eq:pde2T}) on the unit square with periodic boundary conditions in $x$ and $y$, and using the inverse rate $\gamma^{-1}$ as a time unit. We use an operator splitting scheme, whereby the nonlinear terms are advanced via fixed-step fourth order Runge-Kutta, and the diffusion terms by the so-called FTCS scheme, namely a time-explicit
Euler method used jointly with a centered finite difference discretization of the Laplacian terms. Each node of the spatial mesh 
is initialized with random values of the two population fractions, subject to the constraint $C+T<1$.

\begin{figure*}
    \includegraphics[width=0.80\linewidth]{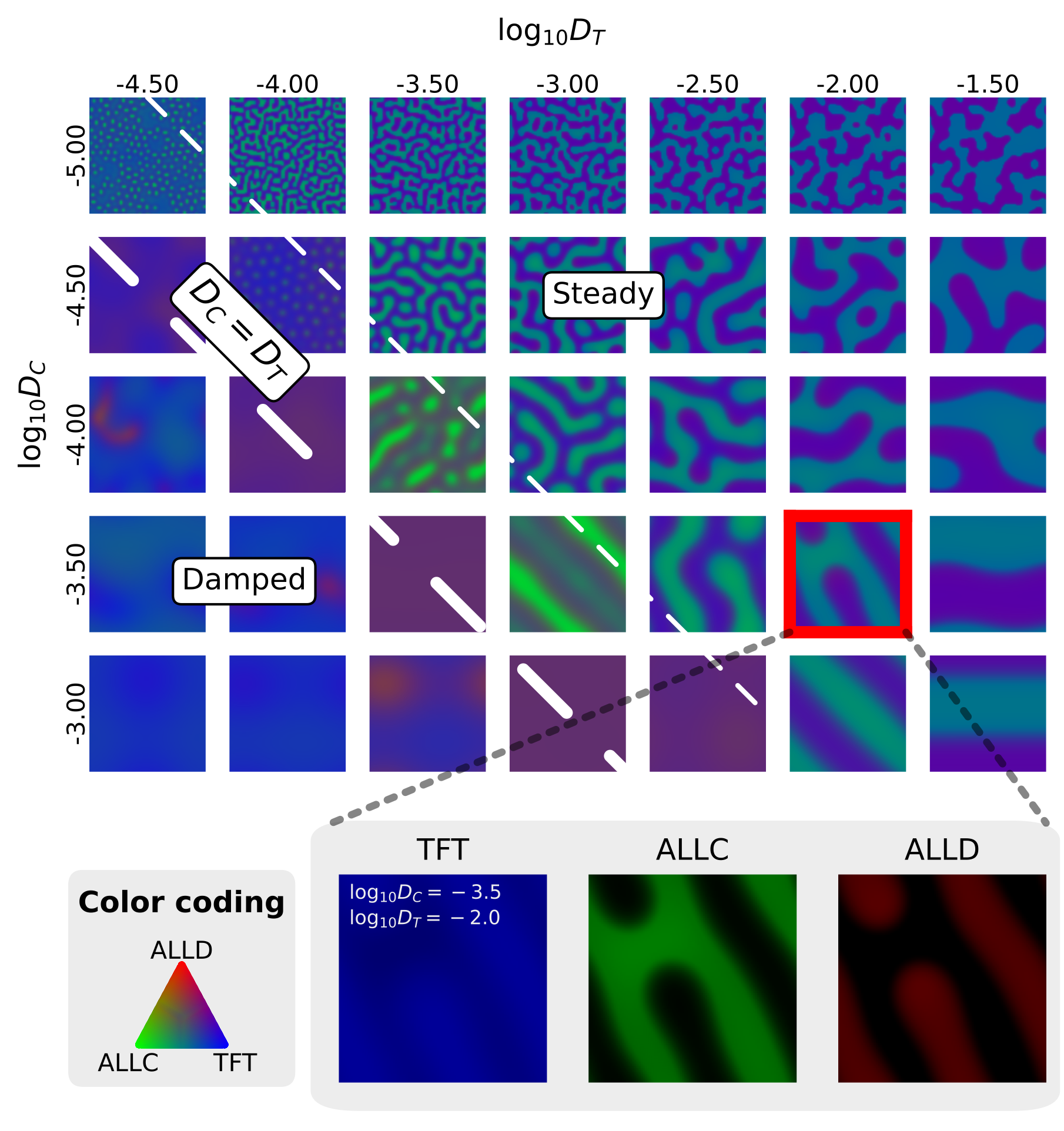}
    \caption{\textbf{Top panels}. Spatial patterns formed in numerical solutions of Eqs.~(\ref{eq:pde2C})--(\ref{eq:pde2T}) in the mean-field model with predation rates $\alpha=0.2$, $\beta=0.5$, and $\gamma=1$, which are equivalent to an error rate of $\log_{10} p\simeq-5$. Solutions are integrated to $t=280\,\gamma^{-1}$, starting from a random initial condition. The diffusion coefficients increases logarithmically moving rightward and downward.
    Above the upper white diagonal ($D_T\geq 4.5D_C$) spatial patterns remain static after nonlinear saturation. When $D_{\rm T} < D_{\rm C}$ (below the main diagonal), transient spatial patterns are progressively smoothed by diffusion. Damped oscillatory patterns materialize in the narrow region between the two diagonals, and no patterns of significant amplitude build up along the main diagonal $D_{\rm T} = D_{\rm C}$. An example of RGB color-coding scheme is shown below the top panels. See also Appendix~\ref{sec:anims} for animated versions of simulations on the main diagonal, and in the steady and damped regimes.}
    \label{fig:ipdODE-snaps}
\end{figure*}

Figure~\ref{fig:ipdODE-snaps} shows a representative set of results, obtained for predation rates $\alpha=0.2$, $\beta=0.5$, and $\gamma=1.0$. The mosaic displays
solutions for increasing $D_T$ (horizontal) and $D_C$ (vertical, and increasing downwards). The thick white dashed line marks the diagonal $D_T=D_C$.
All snapshots are extracted after an elapsed time $t=280\,\gamma^{-1}$, 
and are rendered as a RGB coding of the population fraction,
with blue$\equiv T$ (strategy TFT], green$\equiv C$ (ALLC), and red$\equiv 1-C
-T$ (ALLD). 

The striking spatial patterns located above the white dotted diagonal are
all steady, and set in very early in the evolution; 
Further examination of the temporal evolution of solutions on and below
the $D_T=D_C$ diagonal also reveals global, system-wide oscillations of the population frequencies, with faint spatial patterning
damping out on a timescale decreasing as diffusion becomes more important
(moving towards the bottom right in the mosaic). The solutions
on the first superdiagonal, indicated by the white dotted line, oscillate in time but with a spatially quasi-stationary planform. In all cases the length scale of the spatial patterns, whether steady or oscillating, 
decreases as diffusion diminishes (moving towards the upper left in the mosaic), as one would intuitively expect.

These pattern-forming characteristics, and their dependence on the diffusion coefficients, are similar for other values of the predation rates $\alpha$ and $\beta$, but with the populations fractions of the three strategies (setting the ``color'' in our RGB coding scheme) varying significantly. Steady patterns only
disappear in the extreme regime $\alpha\ll \beta\ll \gamma$, which occurs only at very high error rates in the IPD simulations ($\log_{10} p\geq -1$, viz.~ Fig.~\ref{fig:from-p-to-alpha-beta}). It is well known from studies of pattern formation in classical nonlinear reaction-diﬀusion systems à la Turing that the inhibition variable must diﬀuse faster than the excitation variable for steady patterns to form. Our necessary condition $D_T > D_C$ for stationary patterns similarly implies that the strategy TFT acts here as the inhibitor in our 3-species predator-prey-like system.

\subsection{Stability analysis}

The striking patterns on Fig.~\ref{fig:ipdODE-snaps} can be understood through a linear stability analysis. Working in Cartesian coordinates in two spatial dimensions, the first order linearized versions of 
Eqs.~(\ref{eq:pde2C})--(\ref{eq:pde2T}) about the spatially uniform equilibrium
solutions become: 
\begin{eqnarray}
\label{eq:linpde2C}
\frac{\partial C_1}{\partial t}&=&
\left[{\beta\gamma\over s}\right]C_1
+\left[{(\alpha+\gamma)\beta\over s}\right]T_1+D_C\nabla^2C_1~,\\
\label{eq:linpde2T}
\frac{\partial T_1}{\partial t}&=&
-\left[{(\alpha+\beta)\gamma\over s}\right]C_1
-\left[{\beta\gamma\over s}\right]T_1
+D_T\nabla^2T_1~.
\end{eqnarray}
We are seeking solutions describing exponentially growing but
spatially steady planforms:
\begin{eqnarray}
\label{eq:modal2C}
\begin{pmatrix}
C_1({\bf x},t) \\ T_1({\bf x},t)
\end{pmatrix}
=
\begin{pmatrix}
C_\ast \\ T_\ast
\end{pmatrix}
\exp(i{\bf k\cdot x}+\lambda t)~,
\end{eqnarray}
with $\lambda$ real and positive. Substitution
in Eqs.~(\ref{eq:linpde2C})--(\ref{eq:linpde2T}) yields two coupled
nonlinear algebraic equations, whose solution yields the dispersion relation:
\begin{eqnarray}
\label{eq:dispr}
\lambda^2 &+& \left[ k^2(D_C+D_T) \right] \lambda  \\
&+&\left[D_CD_Tk^4+{\beta\gamma\over s}(D_C-D_T)k^2+{\alpha\beta\gamma\over s}\right]=0,\nonumber
\end{eqnarray}
with $k^2\equiv k_x^2+k_y^2$.
Recalling that $\alpha,\beta,\gamma,D_T$ and $D_C$
are all positive quantities,
a positive $\lambda$ demands that the positive root of this quadratic polynomial be retained,
and for the later to be purely real further requires
\begin{equation}
\label{eq:c}
D_CD_Tk^4
+{\beta\gamma\over s}(D_C-D_T)k^2
+{\alpha\beta\gamma\over s}< 0~.
\end{equation}
This is a convex quadratic polynomial in $k^2$, with its extremum at
\begin{equation}
\label{eq:kmax}
k^2_{\rm max}=
{\beta\gamma(D_T-D_C)\over 2sD_CD_T}~.
\end{equation}
This corresponds to the wavenumber of the most rapidly growing mode.
Note that the requirement that $k_{\rm max}$ be a real quantity demands
$D_T > D_C$. Substituting $k_{\rm max}$ back into Eq.~(\ref{eq:c}) leads to the further constraint:
\begin{equation}
d^2-2(1+A)d+1 > 0~,
\end{equation}
where we have defined $d=D_T/D_C$ and $A=2\alpha s/\beta\gamma$. The positive root
of this quadratic polynomial yields the minimal diffusion coefficient ratio $d_m$ above which the formation of steady spatial pattern is possible;  for the predation rates used to generate Fig.~\ref{fig:ipdODE-snaps}, $d_m=4.5$ (dotted diagonal).

The dispersion relation (\ref{eq:dispr}) also allows oscillations ($\lambda$ picking up an imaginary part) provided its
discriminant is negative:
\begin{equation}
(D_C-D_T)^2k^4-{4\beta\gamma\over s}(D_C-D_T)k^2-{4\alpha\beta\gamma\over s}<0~.
\end{equation}
This is a quadratic in $k^2$, with roots:
\begin{equation}
\label{eq:osclroots}
k^2(\pm) = {2\beta\gamma\over s(D_C-D_T)}
\left[ 1\pm \sqrt{ 1+{\alpha s\over \beta\gamma} } \right].
\end{equation}
Either positive $k^2(+)$ or $k^2(-)$ is possible, according to the sign of $D_C-D_T$.
However, these oscillations are always damped, because the real part of $\lambda$ is now
\begin{equation}
\label{eq:damping}
{\rm Re}(\lambda)=-\frac{1}{2}(D_C+D_T)k^2~.
\end{equation}
Damping is slowest for the largest length scales (small $k$), and
weak diffusion (small $D_C+D_T$), as expected for a diffusively damped oscillation. 

Note that the $k\to 0$ limit brings us back to uniform oscillations of the population fractions, as in \S\ref{ssec:3species}. The oscillation
frequency $\omega\equiv{\rm Im}(\lambda)=\sqrt{\alpha\beta\gamma/s}$ is indeed recovered by setting $k=0$ in the dispersion relation (\ref{eq:dispr}).
Such global oscillations are not expected to  persist
in the nonlinear regime when $D_T/D_C>d(+)$, because they then become overwhelmed by the exponentially growing spatially stationary planform.

Overall, the linear stability analysis thus recovers quite well the pattern-forming characteristics obtained from numerical solutions 
of the nonlinear system (\ref{eq:pde2C})--(\ref{eq:pde2T}). We now turn to the pattern-forming behavior of the spatial IPD model of \S\ref{sec:IPD}, upon the introduction of diffusion.

\section{Diffusion in the cellular automaton model}
\label{sec:diffusion-discrete}

\subsection{Pattern formation}
\label{ssec:pattern-formation}

Our next step is to investigate whether the structures identified in the mean-field model (c.f.,~Fig.~\ref{fig:ipdODE-snaps}) can also arise at the microscale. Exploring the effect of diffusion in the CA model can be carried out using diffusion by interchange: for each iteration $t_i$ of the model, strategies perform a stochastic random walk on the lattice, exchanging place with another strategy in their von Neumann neighborhood (i.e., the four players in each cardinal direction) which is the lattice-based equivalent process of continuous classical linear diffusion (see, e.g.,~\citep{vallis_atmospheric_2017}). While the CA score-based update rules remain purely deterministic, the model
now incorporates stochastic elements in both game playing errors and through interchange diffusion.

\begin{figure*}
    \includegraphics[width=0.80\linewidth]{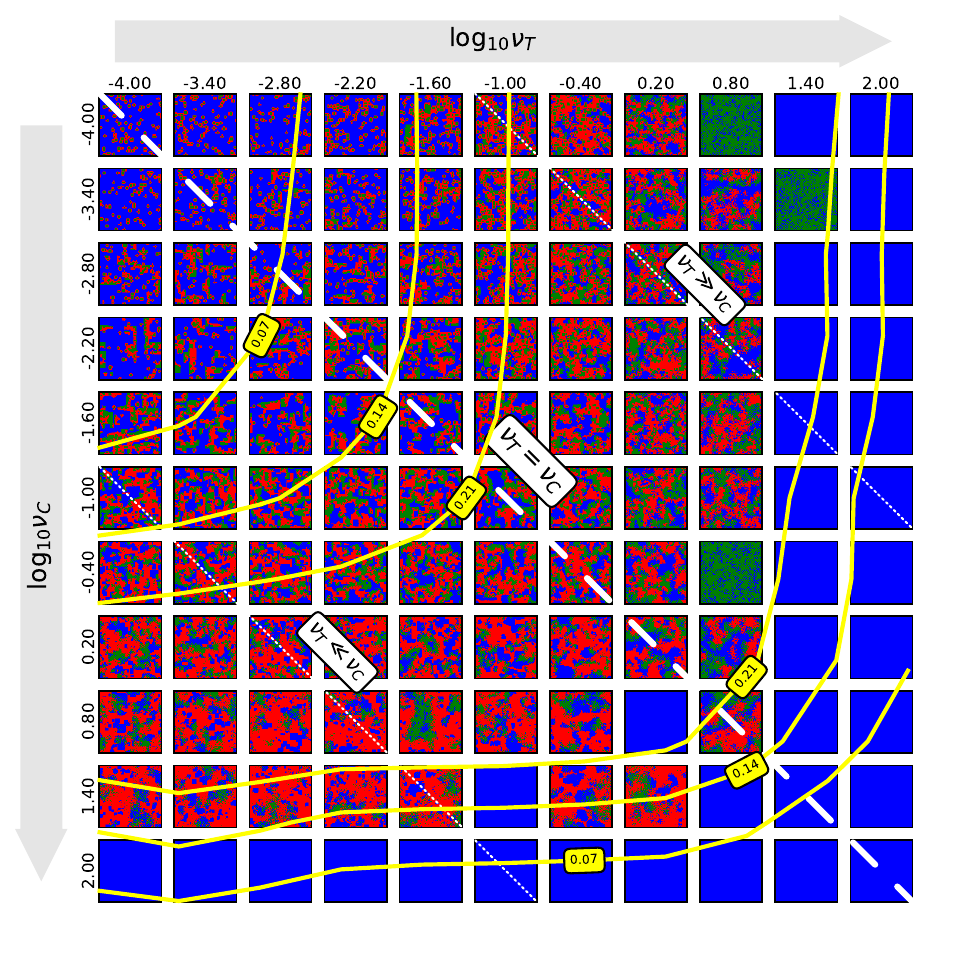}
    \caption{Final lattice state for simulations of the spatial IPD using the cellular automaton model, which includes diffusion and an error rate set at $\log_{10} p=-2$. Simulations were carried out over $T=500$ iterations on a lattice of size $L=128$, with games of $M=2000$ moves. The horizontal axis shows logarithmic variations of TFT’s interchange coefficient while the vertical axis shows that of ALLC players, both ranging from $\log_{10} \nu=-4$ (negligible diffusion) to $\log_{10} \nu=2$ (close to the well-mixed limit). The yellow contour lines shows the degree of non-stationarity, where values approaching zero indicate simulations that become stationary over time whereas higher values those remaining non-stationary. In the upper-left region, ALLC players form stable cooperative clusters surrounded by ALLD strategies (see,~e.g.,~Fig.~\ref{fig:lipd-snaps}c). See also Appendix~\ref{sec:anims} for animated versions of systems on the main ($\nu_T=\nu_C$), upper ($\nu_T \gg \nu_C$) and lower ($\nu_T \ll \nu_C$) diagonals.}
    \label{fig:lipd-spirals}
\end{figure*}

We analyzed the parameter space of the CA model under variation of interchange frequency specified for individual species: Figure~\ref{fig:lipd-spirals} shows the final lattice state for CA simulations where each player on the grid has an interchange frequency $\nu$. The horizontal axis represents the interchange frequency defined for TFT ($\nu_T$), while the vertical axis displays the frequency for ALLC ($\nu_C$). No distinctive behaviors were observed in the CA model for regions of the parameter space where the interchange frequency for ALLD ($\nu_D$) was independently varied.

The interchange frequencies progress logarithmically in both cases where the minimum corresponds to negligible diffusion ($\log_{10} \nu=-4$), where each player will be switching place with one of its neighbor once every $10^{4}$ iterations. Put differently, in this regime where diffusion is minimal, one player will change place every iteration on a $128\times128$ lattice on average. Maximal values shown for the interchange frequencies approach the limit where the system is well-mixed ($\log_{10} \nu=2$), as each player will be switching place $10^2$ times at each iteration.

As both interchange coefficients are increased, (i.e., by moving rightward or downward) larger structures are revealed, whose characteristic length approaches that of the whole lattice. As interchange frequencies increase further, the system transitions to an absorbing state where TFT occupies the whole lattice. Generally, increasing $\nu$ results in the formation of patterns at larger spatial scales, however moving along the $\nu_T \gg \nu_C$ and $\nu_T \ll \nu_C$ diagonals do not reveal symmetrical structures. Moreover, final population fractions between these two cases show significant differences (see discussion in \S\ref{ssec:dom-strat}).

Thus defined, diffusion by interchange necessarily implies coupled diffusion frequencies, as interchange occurs with frequency $\nu_i$ between a player belonging to the species $i$ and another random player $j$ whose strategy can be different. This coupling of the diffusion frequencies between species also implies that the effective $\nu_T$ and $\nu_C$ depend on population frequencies for all three species. Since population frequencies also vary in time, interchange frequencies are also affected when the system enters absorbing states (e.g., when a strategy undergoes stochastic extinction).\\

\subsection{Spatiotemporal variability}

Over time, simulations in the CA model can either stabilize or continue to display diverse degrees of spatiotemporal variability. Letting $\mathbf{G}^t$ be the matrix representing strategies on the lattice at iteration $t$, and $g_{ij}^t$ its values, then the degree of non-stationarity $n$ at time $t$ of a CA simulation corresponds to the number of sites on the lattice where strategy replacement has taken place since iteration $t-1$,

\begin{equation}
    n^t=\frac{1}{L^2} \sum_{\mathbf{G}^t}\Big[1-\delta\left(g_{i j^{\prime}}^t, g_{i j}^{t-1}\right)\Big], \label{eq:nonstationarity}
\end{equation}

\noindent with $\delta(\alpha,\alpha^\prime)$ being the Kronecker delta and $n$ normalized by the size of the lattice. Lower non-stationarity thus implies simulations that become stationary over time while higher non-stationarity indicates simulations that continue to evolve, with $0 \leq n^t \leq 1$. The contour lines on Fig.~\ref{fig:lipd-spirals} show the degree of non-stationarity $n^t$, averaged over statistical ensembles of 10 realizations of the simulation with random initial conditions for each set of parameters.

Varying $\nu_T$ and $\nu_C$ defines several regimes across which system behavior ranges from the formation of small-scale clusters in static equilibrium (top-left region, Fig.~\ref{fig:lipd-spirals}) to structures whose characteristic length scale are of the same order of magnitude as the whole lattice (regions bordering blue panels) which are highly dynamic. Increasing further the interchange frequencies ultimately gives way to a sudden transition towards absorbing states where TFT fills the whole lattice (rightmost and bottom panels) after which the simulation becomes trivially stationary.

\subsection{Equivalence of the models}
\label{ssec:relationship}

We now examine how pattern formation correlates between the mean-field model and the cellular automaton model incorporating diffusion. Inspection of Fig.~\ref{fig:ipdODE-snaps} and Fig.~\ref{fig:lipd-spirals} suggests that moving along both diagonal axes towards increased diffusion results in spatial patterns with increasing characteristic scale. Here, we employ 2D Fourier transforms to formalize this relationship: by calculating the power spectra of the lattices resulting from the execution of simulations with both models, we can quantitatively assess how this increase in wavelength is related to the increase in diffusion.

\begin{figure*}
    \includegraphics[width=0.80\linewidth]{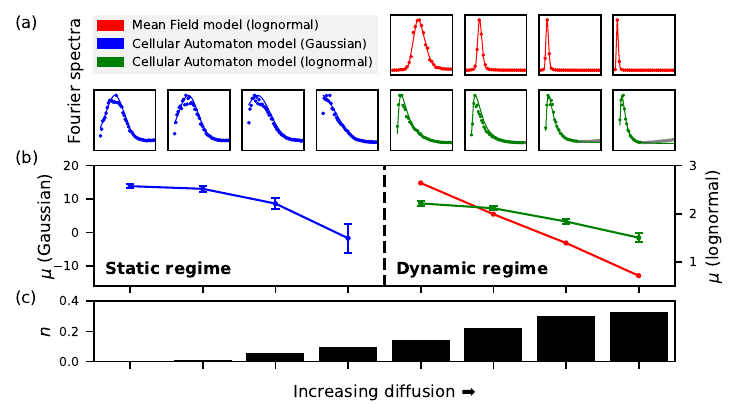}
    \caption{Fourier analysis was performed on the lattices for both the MF and CA models, using statistical ensembles of ten simulations conducted with identical parameters as those used in the simulations on the first super-diagonal (MF model) and the main diagonal (CA model). \textbf{(a)} The Fourier spectra, averaged over the statistical ensemble and best-fit distributions, are plotted against radial wavenumber $k$ ($k^2=k_x^2+k_y^2$) increasing to the right. The spectra for the MF model are depicted in red, while those for the CA model are shown in blue (static regime) and green (dynamic regime). \textbf{(b)} The mean value of the best-fit distributions, averaged over the statistical ensemble, is displayed, with error bars representing the standard deviation. The left side (blue curve) represents the static regime, while the right side (red and green curves) represents the dynamic regime. All three curves demonstrate that for both systems the wavenumber decreases in every regime as diffusion increases. \textbf{(c)} The non-stationarity $n$ is displayed in panel \textbf{(c)}. As diffusion increases, the final latice states gradually transit from static to non-stationary ($n$ increasing).}
    \label{fig:fourier}
\end{figure*}

Utilizing this methodology, we have analyzed the spectra for simulations with parameters identical to the first upper-diagonal of the MF model (Fig.~\ref{fig:ipdODE-snaps}) and the main diagonal of the CA model (Fig.~\ref{fig:lipd-spirals}). The 2D Fourier transform of these spectra, averaged over statistical ensembles of 10 independent simulations of the models, and collapsed on a 1-dimensional axis, are displayed in Fig.~\ref{fig:fourier}a, with the horizontal axis indicating the wavenumber, excluding the simulations where TFT invades the entire lattice in the CA model (since the Fourier spectra would be zero). The spectra for both the MF model (displayed in red) and the dynamic regime of the CA model where simulations remain non-stationary (displayed in green) is well fit by a lognormal distribution, while spectra for the static regime of the CA model are found to align with Gaussian distributions (spectra displayed in blue). Parameters for the best-fit distribution are represented in Fig.~\ref{fig:fourier}b: the mean $\mu$ of the distributions averaged over the statistical ensemble are shown, with error bars indicating the standard deviation over the ensemble.

In Fig.~\ref{fig:fourier}c, we present the non-stationarity, computed using Equation~\eqref{eq:nonstationarity}, at the final iteration of the model. This is averaged over the statistical ensembles of each parameter set, effectively distinguishing between the static regime (where simulations eventually become stationary) and the dynamical regime (where simulations remain non-stationary).

This analysis of the underlying trends in pattern formation across both models demonstrates that the behavior of the MF model reflects that of both the static and dynamic regimes of the CA model: an increase in diffusion results in a decrease of the wavenumber of the patterns, which mirrors what could be inferred by examining Fig.~\ref{fig:ipdODE-snaps} and Fig.~\ref{fig:lipd-spirals} as non-stationarity increases.

\subsection{Dominance of strategies}
\label{ssec:dom-strat}

Increasing either diffusion frequency ($\nu_T$ or $\nu_C$), which corresponds to moving rightward or downward in Fig.~\ref{fig:lipd-spirals}, triggers regime changes in the CA model’s behavior. Moving along both axis induces the formation of large scale structures before well-mixed regimes take over (rightmost and bottom panels in Fig.~\ref{fig:lipd-spirals}). In these well-mixed regimes, the stochastic extinction of one strategy can occur: unlike in the MF model, where populations are continuous values that can come close to but never actually reach zero, the populations in the CA model are discrete. Consequently, if a strategy’s population reaches zero, that species becomes extinct for the remainder of the simulation, and one of the two remaining strategies will be further eliminated, as outlined in \S\ref{ssec:3species}. Careful inspection moreover reveals that upon moving rightward or downward on Fig.~\ref{fig:lipd-spirals}, the system undergoes configuration changes that are not symmetrical about the main diagonal: in these panels, an increase in $\nu_T$ results in a more rapid rise of cooperator populations (ALLC and TFT), contrasting with increases in $\nu_C$ where populations of parasites (ALLD) otherwise tend to dominate.

\begin{figure}[t!]
    \includegraphics[width=1.00\linewidth]{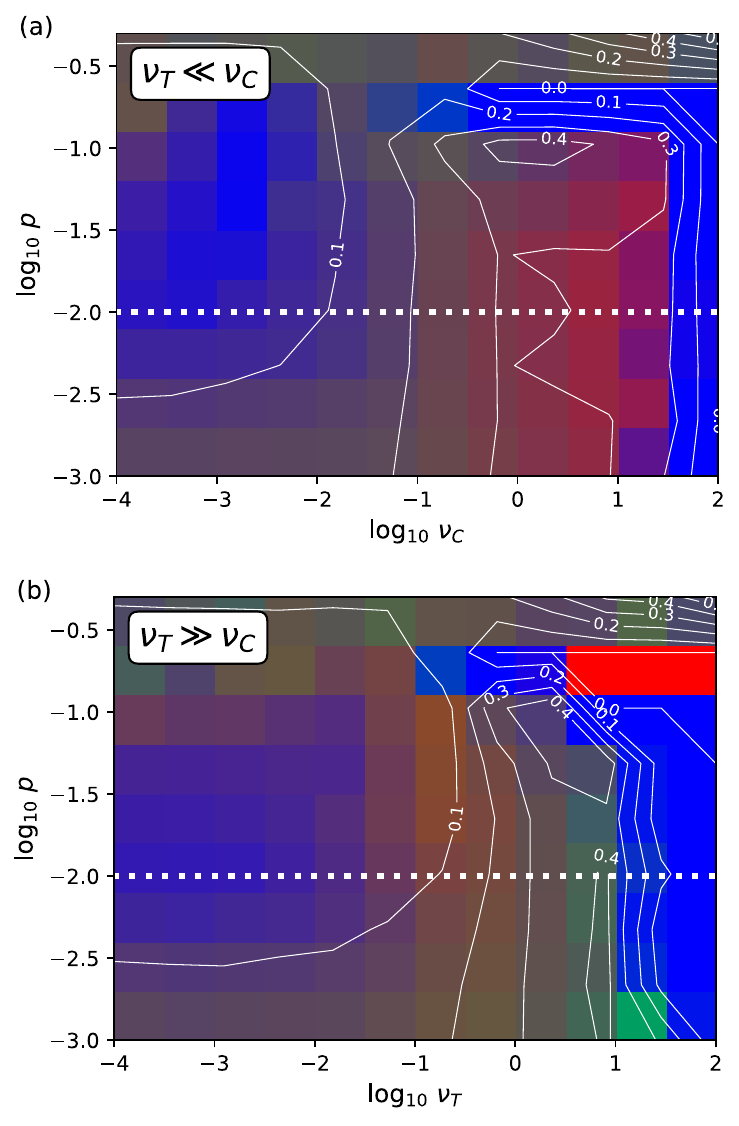}
    \caption{Final population fractions, shown as RGB color coding, for cellular automaton simulations of the spatial IPD which includes diffusion with both varying interchange frequencies and error rates. Population fractions were averaged over statistical ensembles of 10 realizations of the simulation with random initial conditions for each set of parameters. Other parameters such as lattice size, IPD game length and number of iterations are identical to those of simulations shown previously on Fig.~\ref{fig:lipd-spirals}. The horizontal axis represents interchange frequencies, where panel (a) shows the one associated with strategy ALLC (TFT’s interchange frequency being three orders of magnitude smaller) and vice-versa for panel (b). The vertical axis displays the error rate, ranging from a few errors per game ($p=10^{-3}$) to its maximum where game dynamics is random ($p=0.5$). Contour lines show the degree of non-stationarity, where values approaching zero indicate systems that become stationary over time, whereas those with higher values continue to evolve. Moving along the white dotted lines on both panels (a) and (b) amounts to moving across lower and upper diagonals on Fig.~\ref{fig:lipd-spirals} respectively.}
    \label{fig:p-versus-D}
\end{figure}

Investigating further this dissimilarity across error rates, Fig.~\ref{fig:p-versus-D} displays resulting population fractions represented as RGB color coding, each value calculated from statistical ensembles of 10 realizations with random initial conditions, with the vertical axis showing the error rate and the horizontal axis representing the diffusion frequency. In panel (a), the horizontal axis represents $\nu_C$, while $\nu_T$ is three orders of magnitude smaller (i.e., $\nu_T=\nu_C \times 10^{-3}$). In contrast, panel (b) represents $\nu_T$ on its horizontal axis, with $\nu_C$ three orders of magnitude smaller. Slicing horizontally across panels (a) and (b) of Fig.~\ref{fig:p-versus-D} at $\log_{10} p = -2$ corresponds to moving along the lower and upper dotted white diagonals in Fig.~\ref{fig:lipd-spirals}, respectively. Contour lines in Fig.~\ref{fig:p-versus-D} again represent the degree of non-stationarity, as calculated via Equation~\eqref{eq:nonstationarity} and averaged over each 10-member ensemble.

Figure~\ref{fig:p-versus-D} illustrates the contrasting effects that varying either interchange frequency and large-scale structure formation have on resulting population fractions. Panels (a), where $\nu_T \ll \nu_C$, and (b), where $\nu_C \gg \nu_T$, show crucial differences: while at low interchange frequencies $\nu_C$-dominated diffusion favors TFT more explicitly, moving rightward on both panels reveals that cooperative strategies (ALLC and TFT) takes hold for much lower values of the interchange frequency---one exception to this being when the error rate is very high ($\log_{10} p \gtrsim -1.0$). Simply put, the system reaches an absorbing state where TFT invades the lattice much earlier when $\nu_T$ is increased, compared to the scenario where $\nu_C$ is increased. Large-scale pattern formation regions on Fig.~\ref{fig:lipd-spirals} yield higher population fractions for cooperating strategies (ALLC and TFT) in the corresponding regions on panel (b) for $\nu_T$-led diffusion. The implications of TFT-driven diffusion ($\nu_T \gg \nu_C$) and ALLC-driven diffusion ($\nu_T \ll \nu_C$) thus result in very different simulation outcomes. In the latter scenario, parasites gain an advantage, whereas in the former scenario, cooperators are favored at considerably lower interchange frequencies.

Differences between the two regimes shown in Fig.~\ref{fig:p-versus-D} are however muddled due to the coupling of interchange frequencies discussed in \S\ref{ssec:pattern-formation}. System dynamics also differs significantly in parameter regions where the error rate $p$ approaches its maximum, which are shown in the upper regions of both panels on Fig.~\ref{fig:p-versus-D}. Importantly, in the regime where $\nu_T \gg \nu_C$, defectors (ALLD) thrive at high error rates coupled with high interchange frequencies---a behavior which is not observed in the $\nu_T \ll \nu_C$ scenario.

\section{Conclusion}

The emergence, promotion and maintenance of cooperation is an essential feature of various natural phenomena, whether at small or large scale. Our work sheds light on how the presence of diffusive processes in physical systems affects the interplay of strategies competing in the spatial iterated prisoner’s dilemma. We based our analysis on a deterministic cellular automaton model (Section~\ref{sec:IPD}), from which we derived a mean-field model in which we introduced linear diffusion (Section~\ref{sec:mean-field-model}). We further investigated the observed pattern formation through linear stability analysis, then used diffusion by interchange to induce similar dynamics in the deterministic cellular automaton model to characterize its effect on the dominance of strategies (Section~\ref{sec:diffusion-discrete}). Our results indicate that the emergence of patterns whose characteristic length compares with the system’s size---a behavior observed in other classes of systems such as autocatalytic models---significantly influences the populations of strategies playing the IPD and, importantly, is found to favor cooperation.

Our investigations motivates further research in analyzing how diffusion can enable the rise of cooperation in various natural systems. While certain classes of systems such as autocatalytic reaction-diffusion systems are known to display pattern formation, we have shown that similar behavior can emerge in other diffusive systems, with these dynamics fostering cooperative behavior. Further work is needed in order to analyze how the various diffusive processes at work in the context of specific natural systems---such as bacterial colonies or bird flocks---influence coordinated actions and enable the emergence of cooperation between individuals.

\begin{acknowledgments}

Conceptualization, A.C.-R. and P.C.; methodology, A.C.-R. and P.C.; software, A.C.-R. and S.Z.-B.; writing—original draft preparation, A.C.-R., S.Z.-B and P.C.; writing—review and editing, P.C.; funding acquisition, P.C. All authors have read and agreed to the published version of the manuscript.

This research was funded by the Natural Sciences and Engineering Research Council of Canada (grant number RGPIN/05278-2018), the Fonds de recherche Nature et Technologies of Québec (grant number 314488), the Fondation J. Armand Bombardier Excellence Scholarship and the Google Cloud Research Credits Program (grant number 215904839).

\end{acknowledgments}

\appendix

\section{Supplementary Information}
\label{sec:anims}

Animations for simulations presented in this article are available at the following address:\\

\noindent
\href{https://github.com/champagnealexandre/DiffusionPattern}{https://github.com/champagnealexandre/DiffusionPattern}


\begin{thebibliography}{83}%
\makeatletter
\providecommand \@ifxundefined [1]{%
 \@ifx{#1\undefined}
}%
\providecommand \@ifnum [1]{%
 \ifnum #1\expandafter \@firstoftwo
 \else \expandafter \@secondoftwo
 \fi
}%
\providecommand \@ifx [1]{%
 \ifx #1\expandafter \@firstoftwo
 \else \expandafter \@secondoftwo
 \fi
}%
\providecommand \natexlab [1]{#1}%
\providecommand \enquote  [1]{``#1''}%
\providecommand \bibnamefont  [1]{#1}%
\providecommand \bibfnamefont [1]{#1}%
\providecommand \citenamefont [1]{#1}%
\providecommand \href@noop [0]{\@secondoftwo}%
\providecommand \href [0]{\begingroup \@sanitize@url \@href}%
\providecommand \@href[1]{\@@startlink{#1}\@@href}%
\providecommand \@@href[1]{\endgroup#1\@@endlink}%
\providecommand \@sanitize@url [0]{\catcode `\\12\catcode `\$12\catcode
  `\&12\catcode `\#12\catcode `\^12\catcode `\_12\catcode `\%12\relax}%
\providecommand \@@startlink[1]{}%
\providecommand \@@endlink[0]{}%
\providecommand \url  [0]{\begingroup\@sanitize@url \@url }%
\providecommand \@url [1]{\endgroup\@href {#1}{\urlprefix }}%
\providecommand \urlprefix  [0]{URL }%
\providecommand \Eprint [0]{\href }%
\providecommand \doibase [0]{http://dx.doi.org/}%
\providecommand \selectlanguage [0]{\@gobble}%
\providecommand \bibinfo  [0]{\@secondoftwo}%
\providecommand \bibfield  [0]{\@secondoftwo}%
\providecommand \translation [1]{[#1]}%
\providecommand \BibitemOpen [0]{}%
\providecommand \bibitemStop [0]{}%
\providecommand \bibitemNoStop [0]{.\EOS\space}%
\providecommand \EOS [0]{\spacefactor3000\relax}%
\providecommand \BibitemShut  [1]{\csname bibitem#1\endcsname}%
\let\auto@bib@innerbib\@empty
\bibitem [{\citenamefont {Turner}\ and\ \citenamefont
  {Chao}(1999)}]{turner_prisoners_1999}%
  \BibitemOpen
  \bibfield  {author} {\bibinfo {author} {\bibfnamefont {P.~E.}\ \bibnamefont
  {Turner}}\ and\ \bibinfo {author} {\bibfnamefont {L.}~\bibnamefont {Chao}},\
  }\href {\doibase 10.1038/18913} {\bibfield  {journal} {\bibinfo  {journal}
  {Nature}\ }\textbf {\bibinfo {volume} {398}},\ \bibinfo {pages} {441}
  (\bibinfo {year} {1999})}\BibitemShut {NoStop}%
\bibitem [{\citenamefont {Bohl}\ \emph {et~al.}(2014)\citenamefont {Bohl},
  \citenamefont {Hummert}, \citenamefont {Werner}, \citenamefont {Basanta},
  \citenamefont {Deutsch}, \citenamefont {Schuster}, \citenamefont {Theißen},\
  and\ \citenamefont {Schroeter}}]{bohl_evolutionary_2014}%
  \BibitemOpen
  \bibfield  {author} {\bibinfo {author} {\bibfnamefont {K.}~\bibnamefont
  {Bohl}}, \bibinfo {author} {\bibfnamefont {S.}~\bibnamefont {Hummert}},
  \bibinfo {author} {\bibfnamefont {S.}~\bibnamefont {Werner}}, \bibinfo
  {author} {\bibfnamefont {D.}~\bibnamefont {Basanta}}, \bibinfo {author}
  {\bibfnamefont {A.}~\bibnamefont {Deutsch}}, \bibinfo {author} {\bibfnamefont
  {S.}~\bibnamefont {Schuster}}, \bibinfo {author} {\bibfnamefont
  {G.}~\bibnamefont {Theißen}}, \ and\ \bibinfo {author} {\bibfnamefont
  {A.}~\bibnamefont {Schroeter}},\ }\href {\doibase 10.1039/C3MB70601J}
  {\bibfield  {journal} {\bibinfo  {journal} {Molecular BioSystems}\ }\textbf
  {\bibinfo {volume} {10}},\ \bibinfo {pages} {3066} (\bibinfo {year}
  {2014})}\BibitemShut {NoStop}%
\bibitem [{\citenamefont {Hong}\ \emph {et~al.}(1992)\citenamefont {Hong},
  \citenamefont {Feng}, \citenamefont {Rotello},\ and\ \citenamefont
  {Rebek}}]{hong_competition_1992}%
  \BibitemOpen
  \bibfield  {author} {\bibinfo {author} {\bibfnamefont {J.~I.}\ \bibnamefont
  {Hong}}, \bibinfo {author} {\bibfnamefont {Q.}~\bibnamefont {Feng}}, \bibinfo
  {author} {\bibfnamefont {V.}~\bibnamefont {Rotello}}, \ and\ \bibinfo
  {author} {\bibfnamefont {J.}~\bibnamefont {Rebek}},\ }\href {\doibase
  10.1126/science.255.5046.848} {\bibfield  {journal} {\bibinfo  {journal}
  {Science (New York, N.Y.)}\ }\textbf {\bibinfo {volume} {255}},\ \bibinfo
  {pages} {848} (\bibinfo {year} {1992})}\BibitemShut {NoStop}%
\bibitem [{\citenamefont {Sievers}\ and\ \citenamefont {von
  Kiedrowski}(1994)}]{sievers_self-replication_1994}%
  \BibitemOpen
  \bibfield  {author} {\bibinfo {author} {\bibfnamefont {D.}~\bibnamefont
  {Sievers}}\ and\ \bibinfo {author} {\bibfnamefont {G.}~\bibnamefont {von
  Kiedrowski}},\ }\href {\doibase 10.1038/369221a0} {\bibfield  {journal}
  {\bibinfo  {journal} {Nature}\ }\textbf {\bibinfo {volume} {369}},\ \bibinfo
  {pages} {221} (\bibinfo {year} {1994})}\BibitemShut {NoStop}%
\bibitem [{\citenamefont {Lee}\ \emph {et~al.}(1997)\citenamefont {Lee},
  \citenamefont {Severin},\ and\ \citenamefont
  {Ghadiri}}]{lee_autocatalytic_1997}%
  \BibitemOpen
  \bibfield  {author} {\bibinfo {author} {\bibfnamefont {D.~H.}\ \bibnamefont
  {Lee}}, \bibinfo {author} {\bibfnamefont {K.}~\bibnamefont {Severin}}, \ and\
  \bibinfo {author} {\bibfnamefont {M.~R.}\ \bibnamefont {Ghadiri}},\ }\href
  {\doibase 10.1016/s1367-5931(97)80043-9} {\bibfield  {journal} {\bibinfo
  {journal} {Current Opinion in Chemical Biology}\ }\textbf {\bibinfo {volume}
  {1}},\ \bibinfo {pages} {491} (\bibinfo {year} {1997})}\BibitemShut {NoStop}%
\bibitem [{\citenamefont {Dugatkin}(1997)}]{dugatkin_cooperation_1997}%
  \BibitemOpen
  \bibfield  {author} {\bibinfo {author} {\bibfnamefont {L.~A.}\ \bibnamefont
  {Dugatkin}},\ }\href@noop {} {\emph {\bibinfo {title} {Cooperation among
  animals: an evolutionary perspective}}}\ (\bibinfo  {publisher} {Oxford
  University Press},\ \bibinfo {address} {New York},\ \bibinfo {year}
  {1997})\BibitemShut {NoStop}%
\bibitem [{\citenamefont {Dugatkin}(1988)}]{dugatkin_guppies_1988}%
  \BibitemOpen
  \bibfield  {author} {\bibinfo {author} {\bibfnamefont {L.~A.}\ \bibnamefont
  {Dugatkin}},\ }\href {\doibase 10.1007/BF00303714} {\bibfield  {journal}
  {\bibinfo  {journal} {Behavioral Ecology and Sociobiology}\ }\textbf
  {\bibinfo {volume} {23}},\ \bibinfo {pages} {395} (\bibinfo {year}
  {1988})}\BibitemShut {NoStop}%
\bibitem [{\citenamefont {Heinsohn}\ and\ \citenamefont
  {Packer}(1995)}]{heinsohn_complex_1995}%
  \BibitemOpen
  \bibfield  {author} {\bibinfo {author} {\bibfnamefont {R.}~\bibnamefont
  {Heinsohn}}\ and\ \bibinfo {author} {\bibfnamefont {C.}~\bibnamefont
  {Packer}},\ }\href {\doibase 10.1126/science.7652573} {\bibfield  {journal}
  {\bibinfo  {journal} {Science}\ }\textbf {\bibinfo {volume} {269}},\ \bibinfo
  {pages} {1260} (\bibinfo {year} {1995})}\BibitemShut {NoStop}%
\bibitem [{\citenamefont {Kay}\ \emph {et~al.}(2020)\citenamefont {Kay},
  \citenamefont {Keller},\ and\ \citenamefont {Lehmann}}]{kay_evolution_2020}%
  \BibitemOpen
  \bibfield  {author} {\bibinfo {author} {\bibfnamefont {T.}~\bibnamefont
  {Kay}}, \bibinfo {author} {\bibfnamefont {L.}~\bibnamefont {Keller}}, \ and\
  \bibinfo {author} {\bibfnamefont {L.}~\bibnamefont {Lehmann}},\ }\href
  {\doibase 10.1073/pnas.2013596117} {\bibfield  {journal} {\bibinfo  {journal}
  {Proceedings of the National Academy of Sciences}\ }\textbf {\bibinfo
  {volume} {117}},\ \bibinfo {pages} {28894} (\bibinfo {year}
  {2020})}\BibitemShut {NoStop}%
\bibitem [{\citenamefont {Lombardo}(1985)}]{lombardo_mutual_1985}%
  \BibitemOpen
  \bibfield  {author} {\bibinfo {author} {\bibfnamefont {M.~P.}\ \bibnamefont
  {Lombardo}},\ }\href {\doibase 10.1126/science.227.4692.1363} {\bibfield
  {journal} {\bibinfo  {journal} {Science}\ }\textbf {\bibinfo {volume}
  {227}},\ \bibinfo {pages} {1363} (\bibinfo {year} {1985})}\BibitemShut
  {NoStop}%
\bibitem [{\citenamefont {Milinski}(1987)}]{milinski_tit_1987}%
  \BibitemOpen
  \bibfield  {author} {\bibinfo {author} {\bibfnamefont {M.}~\bibnamefont
  {Milinski}},\ }\href {\doibase 10.1038/325433a0} {\bibfield  {journal}
  {\bibinfo  {journal} {Nature}\ }\textbf {\bibinfo {volume} {325}},\ \bibinfo
  {pages} {433} (\bibinfo {year} {1987})}\BibitemShut {NoStop}%
\bibitem [{\citenamefont {Sagan}(1967)}]{sagan_origin_1967}%
  \BibitemOpen
  \bibfield  {author} {\bibinfo {author} {\bibfnamefont {L.}~\bibnamefont
  {Sagan}},\ }\href {\doibase 10.1016/0022-5193(67)90079-3} {\bibfield
  {journal} {\bibinfo  {journal} {Journal of theoretical biology}\ }\textbf
  {\bibinfo {volume} {14}},\ \bibinfo {pages} {225} (\bibinfo {year}
  {1967})}\BibitemShut {NoStop}%
\bibitem [{\citenamefont {Margulis}(1981)}]{margulis_symbiosis_1981}%
  \BibitemOpen
  \bibfield  {author} {\bibinfo {author} {\bibfnamefont {L.}~\bibnamefont
  {Margulis}},\ }\href@noop {} {\emph {\bibinfo {title} {Symbiosis in {Cell}
  {Evolution}: {Life} and {Its} {Environment} on the {Early} {Earth}}}}\
  (\bibinfo  {publisher} {W.H. Freeman \& Company},\ \bibinfo {address} {San
  Francisco},\ \bibinfo {year} {1981})\BibitemShut {NoStop}%
\bibitem [{\citenamefont {Kauffman}(1986)}]{kauffman_autocatalytic_1986}%
  \BibitemOpen
  \bibfield  {author} {\bibinfo {author} {\bibfnamefont {S.~A.}\ \bibnamefont
  {Kauffman}},\ }\href {\doibase 10.1016/S0022-5193(86)80047-9} {\bibfield
  {journal} {\bibinfo  {journal} {Journal of Theoretical Biology}\ }\textbf
  {\bibinfo {volume} {119}},\ \bibinfo {pages} {1} (\bibinfo {year}
  {1986})}\BibitemShut {NoStop}%
\bibitem [{\citenamefont {Hordijk}\ and\ \citenamefont
  {Steel}(2004)}]{hordijk_detecting_2004}%
  \BibitemOpen
  \bibfield  {author} {\bibinfo {author} {\bibfnamefont {W.}~\bibnamefont
  {Hordijk}}\ and\ \bibinfo {author} {\bibfnamefont {M.}~\bibnamefont
  {Steel}},\ }\href {\doibase 10.1016/j.jtbi.2003.11.020} {\bibfield  {journal}
  {\bibinfo  {journal} {Journal of Theoretical Biology}\ }\textbf {\bibinfo
  {volume} {227}},\ \bibinfo {pages} {451} (\bibinfo {year}
  {2004})}\BibitemShut {NoStop}%
\bibitem [{\citenamefont {Hordijk}(2019)}]{hordijk_history_2019}%
  \BibitemOpen
  \bibfield  {author} {\bibinfo {author} {\bibfnamefont {W.}~\bibnamefont
  {Hordijk}},\ }\href {\doibase 10.1007/s13752-019-00330-w} {\bibfield
  {journal} {\bibinfo  {journal} {Biological Theory}\ }\textbf {\bibinfo
  {volume} {14}},\ \bibinfo {pages} {224} (\bibinfo {year} {2019})}\BibitemShut
  {NoStop}%
\bibitem [{\citenamefont {Queller}(1997)}]{queller_cooperators_1997}%
  \BibitemOpen
  \bibfield  {author} {\bibinfo {author} {\bibfnamefont {D.~C.}\ \bibnamefont
  {Queller}},\ }\href {\doibase 10.1086/419766} {\bibfield  {journal} {\bibinfo
   {journal} {The Quarterly Review of Biology}\ }\textbf {\bibinfo {volume}
  {72}},\ \bibinfo {pages} {184} (\bibinfo {year} {1997})}\BibitemShut
  {NoStop}%
\bibitem [{\citenamefont {Szathmáry}(2015)}]{szathmary_toward_2015}%
  \BibitemOpen
  \bibfield  {author} {\bibinfo {author} {\bibfnamefont {E.}~\bibnamefont
  {Szathmáry}},\ }\href {\doibase 10.1073/pnas.1421398112} {\bibfield
  {journal} {\bibinfo  {journal} {Proceedings of the National Academy of
  Sciences}\ }\textbf {\bibinfo {volume} {112}},\ \bibinfo {pages} {10104}
  (\bibinfo {year} {2015})}\BibitemShut {NoStop}%
\bibitem [{\citenamefont {Massey}\ and\ \citenamefont
  {Mishra}(2018)}]{massey_origin_2018}%
  \BibitemOpen
  \bibfield  {author} {\bibinfo {author} {\bibfnamefont {S.~E.}\ \bibnamefont
  {Massey}}\ and\ \bibinfo {author} {\bibfnamefont {B.}~\bibnamefont
  {Mishra}},\ }\href {\doibase 10.1098/rsif.2018.0429} {\bibfield  {journal}
  {\bibinfo  {journal} {Journal of The Royal Society Interface}\ }\textbf
  {\bibinfo {volume} {15}},\ \bibinfo {pages} {20180429} (\bibinfo {year}
  {2018})}\BibitemShut {NoStop}%
\bibitem [{\citenamefont {Colman}\ and\ \citenamefont
  {Colman}(2013)}]{colman_game_2013}%
  \BibitemOpen
  \bibfield  {author} {\bibinfo {author} {\bibfnamefont {A.~M.}\ \bibnamefont
  {Colman}}\ and\ \bibinfo {author} {\bibfnamefont {A.~M.}\ \bibnamefont
  {Colman}},\ }\href
  {https://nls.ldls.org.uk/welcome.html?ark:/81055/vdc_100060363583.0x000001}
  {\emph {\bibinfo {title} {Game theory and its applications in the social and
  biological sciences}}}\ (\bibinfo {year} {2013})\BibitemShut {NoStop}%
\bibitem [{\citenamefont {Myerson}(2013)}]{myerson_game_2013}%
  \BibitemOpen
  \bibfield  {author} {\bibinfo {author} {\bibfnamefont {R.~B.}\ \bibnamefont
  {Myerson}},\ }\href@noop {} {\emph {\bibinfo {title} {Game {Theory}
  {Analysis} of {Conflict}}}}\ (\bibinfo  {publisher} {Harvard University
  Press},\ \bibinfo {address} {Cumberland},\ \bibinfo {year}
  {2013})\BibitemShut {NoStop}%
\bibitem [{\citenamefont {Watson}(2016)}]{watson_strategy_2016}%
  \BibitemOpen
  \bibfield  {author} {\bibinfo {author} {\bibfnamefont {J.}~\bibnamefont
  {Watson}},\ }\href@noop {} {\emph {\bibinfo {title} {Strategy: an
  introduction to game theory}}}\ (\bibinfo  {publisher} {Norton \& Company},\
  \bibinfo {address} {New York; London [etc.},\ \bibinfo {year}
  {2016})\BibitemShut {NoStop}%
\bibitem [{\citenamefont {Sigmund}(1995)}]{sigmund_games_1995}%
  \BibitemOpen
  \bibfield  {author} {\bibinfo {author} {\bibfnamefont {K.}~\bibnamefont
  {Sigmund}},\ }\href@noop {} {\emph {\bibinfo {title} {Games of life:
  explorations in ecology, evolution, and behaviour}}}\ (\bibinfo {year}
  {1995})\BibitemShut {NoStop}%
\bibitem [{\citenamefont {Keinan}\ \emph {et~al.}(2006)\citenamefont {Keinan},
  \citenamefont {Sandbank}, \citenamefont {Hilgetag}, \citenamefont
  {Meilijson},\ and\ \citenamefont {Ruppin}}]{keinan_axiomatic_2006}%
  \BibitemOpen
  \bibfield  {author} {\bibinfo {author} {\bibfnamefont {A.}~\bibnamefont
  {Keinan}}, \bibinfo {author} {\bibfnamefont {B.}~\bibnamefont {Sandbank}},
  \bibinfo {author} {\bibfnamefont {C.~C.}\ \bibnamefont {Hilgetag}}, \bibinfo
  {author} {\bibfnamefont {I.}~\bibnamefont {Meilijson}}, \ and\ \bibinfo
  {author} {\bibfnamefont {E.}~\bibnamefont {Ruppin}},\ }\href {\doibase
  10.1162/artl.2006.12.3.333} {\bibfield  {journal} {\bibinfo  {journal}
  {Artificial Life}\ }\textbf {\bibinfo {volume} {12}},\ \bibinfo {pages} {333}
  (\bibinfo {year} {2006})}\BibitemShut {NoStop}%
\bibitem [{\citenamefont {Schuster}\ \emph {et~al.}(2008)\citenamefont
  {Schuster}, \citenamefont {Kreft}, \citenamefont {Schroeter},\ and\
  \citenamefont {Pfeiffer}}]{schuster_use_2008}%
  \BibitemOpen
  \bibfield  {author} {\bibinfo {author} {\bibfnamefont {S.}~\bibnamefont
  {Schuster}}, \bibinfo {author} {\bibfnamefont {J.-U.}\ \bibnamefont {Kreft}},
  \bibinfo {author} {\bibfnamefont {A.}~\bibnamefont {Schroeter}}, \ and\
  \bibinfo {author} {\bibfnamefont {T.}~\bibnamefont {Pfeiffer}},\ }\href
  {\doibase 10.1007/s10867-008-9101-4} {\bibfield  {journal} {\bibinfo
  {journal} {Journal of Biological Physics}\ }\textbf {\bibinfo {volume}
  {34}},\ \bibinfo {pages} {1} (\bibinfo {year} {2008})}\BibitemShut {NoStop}%
\bibitem [{\citenamefont {Fudenberg}\ and\ \citenamefont
  {Tirole}(1991)}]{fudenberg_game_1991}%
  \BibitemOpen
  \bibfield  {author} {\bibinfo {author} {\bibfnamefont {D.}~\bibnamefont
  {Fudenberg}}\ and\ \bibinfo {author} {\bibfnamefont {J.}~\bibnamefont
  {Tirole}},\ }\href@noop {} {\emph {\bibinfo {title} {Game theory}}}\
  (\bibinfo  {publisher} {MIT Press},\ \bibinfo {address} {Cambridge, Mass.},\
  \bibinfo {year} {1991})\BibitemShut {NoStop}%
\bibitem [{\citenamefont {Nowak}(2006)}]{nowak_evolutionary_2006}%
  \BibitemOpen
  \bibfield  {author} {\bibinfo {author} {\bibfnamefont {M.~A.}\ \bibnamefont
  {Nowak}},\ }\href@noop {} {\emph {\bibinfo {title} {Evolutionary dynamics:
  exploring the equations of life}}}\ (\bibinfo  {publisher} {Belknap Press of
  Harvard University Press},\ \bibinfo {address} {Cambridge, Mass},\ \bibinfo
  {year} {2006})\BibitemShut {NoStop}%
\bibitem [{\citenamefont {Trivers}(1985)}]{trivers_social_1985}%
  \BibitemOpen
  \bibfield  {author} {\bibinfo {author} {\bibfnamefont {R.}~\bibnamefont
  {Trivers}},\ }\href@noop {} {\emph {\bibinfo {title} {Social {Evolution}}}}\
  (\bibinfo  {publisher} {Benjamin-Cummings Pub Co},\ \bibinfo {address} {Menlo
  Park, Calif},\ \bibinfo {year} {1985})\BibitemShut {NoStop}%
\bibitem [{\citenamefont {Delahaye}\ and\ \citenamefont
  {Mathieu}(1995)}]{delahaye_complex_1995}%
  \BibitemOpen
  \bibfield  {author} {\bibinfo {author} {\bibfnamefont {J.-P.}\ \bibnamefont
  {Delahaye}}\ and\ \bibinfo {author} {\bibfnamefont {P.}~\bibnamefont
  {Mathieu}},\ }\href@noop {} {\enquote {\bibinfo {title} {Complex {Strategies}
  in the {Iterated} {Prisoner}'s {Dilemma}},}\ } (\bibinfo {year}
  {1995})\BibitemShut {NoStop}%
\bibitem [{\citenamefont {Jurišić}\ \emph {et~al.}(2012)\citenamefont
  {Jurišić}, \citenamefont {Kermek},\ and\ \citenamefont
  {Konecki}}]{jurisic_review_2012}%
  \BibitemOpen
  \bibfield  {author} {\bibinfo {author} {\bibfnamefont {M.}~\bibnamefont
  {Jurišić}}, \bibinfo {author} {\bibfnamefont {D.}~\bibnamefont {Kermek}}, \
  and\ \bibinfo {author} {\bibfnamefont {M.}~\bibnamefont {Konecki}},\ }in\
  \href@noop {} {\emph {\bibinfo {booktitle} {2012 {Proceedings} of the 35th
  {International} {Convention} {MIPRO}}}}\ (\bibinfo {year} {2012})\ pp.\
  \bibinfo {pages} {1093--1097}\BibitemShut {NoStop}%
\bibitem [{\citenamefont {Haider}(2005)}]{haider_using_2005}%
  \BibitemOpen
  \bibfield  {author} {\bibinfo {author} {\bibfnamefont {A.}~\bibnamefont
  {Haider}},\ }\href {https://mpra.ub.uni-muenchen.de/28574/} {\enquote
  {\bibinfo {title} {Using {Genetic} {Algorithms} to {Develop} {Strategies} for
  the {Prisoners} {Dilemma}},}\ } (\bibinfo {year} {2005})\BibitemShut
  {NoStop}%
\bibitem [{\citenamefont {Sandholm}\ and\ \citenamefont
  {Crites}(1996)}]{sandholm_multiagent_1996}%
  \BibitemOpen
  \bibfield  {author} {\bibinfo {author} {\bibfnamefont {T.~W.}\ \bibnamefont
  {Sandholm}}\ and\ \bibinfo {author} {\bibfnamefont {R.~H.}\ \bibnamefont
  {Crites}},\ }\href {\doibase 10.1016/0303-2647(95)01551-5} {\bibfield
  {journal} {\bibinfo  {journal} {Biosystems}\ }\textbf {\bibinfo {volume}
  {37}},\ \bibinfo {pages} {147} (\bibinfo {year} {1996})}\BibitemShut
  {NoStop}%
\bibitem [{\citenamefont {Axelrod}\ and\ \citenamefont
  {Hamilton}(1981)}]{axelrod_evolution_1981}%
  \BibitemOpen
  \bibfield  {author} {\bibinfo {author} {\bibfnamefont {R.}~\bibnamefont
  {Axelrod}}\ and\ \bibinfo {author} {\bibfnamefont {W.~D.}\ \bibnamefont
  {Hamilton}},\ }\href {\doibase 10.1126/science.7466396} {\bibfield  {journal}
  {\bibinfo  {journal} {Science}\ ,\ \bibinfo {pages} {11}} (\bibinfo {year}
  {1981})}\BibitemShut {NoStop}%
\bibitem [{\citenamefont {Axelrod}(1984)}]{axelrod_evolution_1984}%
  \BibitemOpen
  \bibfield  {author} {\bibinfo {author} {\bibfnamefont {R.~M.}\ \bibnamefont
  {Axelrod}},\ }\href@noop {} {\emph {\bibinfo {title} {The evolution of
  cooperation}}},\ \bibinfo {edition} {rev. ed}\ ed.\ (\bibinfo  {publisher}
  {Basic Books},\ \bibinfo {address} {New York},\ \bibinfo {year}
  {1984})\BibitemShut {NoStop}%
\bibitem [{\citenamefont {Nowak}\ and\ \citenamefont
  {Sigmund}(1992)}]{nowak_tit_1992}%
  \BibitemOpen
  \bibfield  {author} {\bibinfo {author} {\bibfnamefont {M.~A.}\ \bibnamefont
  {Nowak}}\ and\ \bibinfo {author} {\bibfnamefont {K.}~\bibnamefont
  {Sigmund}},\ }\href {\doibase 10.1038/355250a0} {\bibfield  {journal}
  {\bibinfo  {journal} {Nature}\ }\textbf {\bibinfo {volume} {355}},\ \bibinfo
  {pages} {250} (\bibinfo {year} {1992})}\BibitemShut {NoStop}%
\bibitem [{\citenamefont {Reif}(2009)}]{reif_fundamentals_2009}%
  \BibitemOpen
  \bibfield  {author} {\bibinfo {author} {\bibfnamefont {F.}~\bibnamefont
  {Reif}},\ }\href@noop {} {\emph {\bibinfo {title} {Fundamentals of
  statistical and thermal physics}}}\ (\bibinfo  {publisher} {Waveland Press},\
  \bibinfo {address} {Long Grove, Illinois},\ \bibinfo {year}
  {2009})\BibitemShut {NoStop}%
\bibitem [{\citenamefont {Szabó}\ and\ \citenamefont
  {Tőke}(1998)}]{szabo_evolutionary_1998}%
  \BibitemOpen
  \bibfield  {author} {\bibinfo {author} {\bibfnamefont {G.}~\bibnamefont
  {Szabó}}\ and\ \bibinfo {author} {\bibfnamefont {C.}~\bibnamefont {Tőke}},\
  }\href {\doibase 10.1103/PhysRevE.58.69} {\bibfield  {journal} {\bibinfo
  {journal} {Physical Review E}\ }\textbf {\bibinfo {volume} {58}},\ \bibinfo
  {pages} {69} (\bibinfo {year} {1998})}\BibitemShut {NoStop}%
\bibitem [{\citenamefont {Lim}\ \emph {et~al.}(2002)\citenamefont {Lim},
  \citenamefont {Chen},\ and\ \citenamefont
  {Jayaprakash}}]{lim_scale-invariant_2002}%
  \BibitemOpen
  \bibfield  {author} {\bibinfo {author} {\bibfnamefont {Y.~F.}\ \bibnamefont
  {Lim}}, \bibinfo {author} {\bibfnamefont {K.}~\bibnamefont {Chen}}, \ and\
  \bibinfo {author} {\bibfnamefont {C.}~\bibnamefont {Jayaprakash}},\ }\href
  {\doibase 10.1103/PhysRevE.65.026134} {\bibfield  {journal} {\bibinfo
  {journal} {Physical Review E}\ }\textbf {\bibinfo {volume} {65}} (\bibinfo
  {year} {2002}),\ 10.1103/PhysRevE.65.026134}\BibitemShut {NoStop}%
\bibitem [{\citenamefont {Szabó}\ \emph {et~al.}(2005)\citenamefont {Szabó},
  \citenamefont {Vukov},\ and\ \citenamefont {Szolnoki}}]{szabo_phase_2005}%
  \BibitemOpen
  \bibfield  {author} {\bibinfo {author} {\bibfnamefont {G.}~\bibnamefont
  {Szabó}}, \bibinfo {author} {\bibfnamefont {J.}~\bibnamefont {Vukov}}, \
  and\ \bibinfo {author} {\bibfnamefont {A.}~\bibnamefont {Szolnoki}},\ }\href
  {\doibase 10.1103/PhysRevE.72.047107} {\bibfield  {journal} {\bibinfo
  {journal} {Physical Review E}\ }\textbf {\bibinfo {volume} {72}},\ \bibinfo
  {pages} {047107} (\bibinfo {year} {2005})}\BibitemShut {NoStop}%
\bibitem [{\citenamefont {Chen}\ \emph {et~al.}(2008)\citenamefont {Chen},
  \citenamefont {Fu},\ and\ \citenamefont {Wang}}]{chen_interaction_2008}%
  \BibitemOpen
  \bibfield  {author} {\bibinfo {author} {\bibfnamefont {X.}~\bibnamefont
  {Chen}}, \bibinfo {author} {\bibfnamefont {F.}~\bibnamefont {Fu}}, \ and\
  \bibinfo {author} {\bibfnamefont {L.}~\bibnamefont {Wang}},\ }\href {\doibase
  10.1103/PhysRevE.78.051120} {\bibfield  {journal} {\bibinfo  {journal}
  {Physical Review E}\ }\textbf {\bibinfo {volume} {78}} (\bibinfo {year}
  {2008}),\ 10.1103/PhysRevE.78.051120}\BibitemShut {NoStop}%
\bibitem [{\citenamefont {Szolnoki}\ \emph {et~al.}(2009)\citenamefont
  {Szolnoki}, \citenamefont {Vukov},\ and\ \citenamefont
  {Szabó}}]{szolnoki_selection_2009}%
  \BibitemOpen
  \bibfield  {author} {\bibinfo {author} {\bibfnamefont {A.}~\bibnamefont
  {Szolnoki}}, \bibinfo {author} {\bibfnamefont {J.}~\bibnamefont {Vukov}}, \
  and\ \bibinfo {author} {\bibfnamefont {G.}~\bibnamefont {Szabó}},\ }\href
  {\doibase 10.1103/PhysRevE.80.056112} {\bibfield  {journal} {\bibinfo
  {journal} {Physical Review E}\ }\textbf {\bibinfo {volume} {80}} (\bibinfo
  {year} {2009}),\ 10.1103/PhysRevE.80.056112}\BibitemShut {NoStop}%
\bibitem [{\citenamefont {Fant}\ \emph {et~al.}(2023)\citenamefont {Fant},
  \citenamefont {Mazzarisi}, \citenamefont {Panizon},\ and\ \citenamefont
  {Grilli}}]{fant_stable_2023}%
  \BibitemOpen
  \bibfield  {author} {\bibinfo {author} {\bibfnamefont {L.}~\bibnamefont
  {Fant}}, \bibinfo {author} {\bibfnamefont {O.}~\bibnamefont {Mazzarisi}},
  \bibinfo {author} {\bibfnamefont {E.}~\bibnamefont {Panizon}}, \ and\
  \bibinfo {author} {\bibfnamefont {J.}~\bibnamefont {Grilli}},\ }\href
  {\doibase 10.1103/PhysRevE.108.L012401} {\bibfield  {journal} {\bibinfo
  {journal} {Physical Review E}\ }\textbf {\bibinfo {volume} {108}},\ \bibinfo
  {pages} {L012401} (\bibinfo {year} {2023})}\BibitemShut {NoStop}%
\bibitem [{\citenamefont {Szabó}\ \emph {et~al.}(2000)\citenamefont {Szabó},
  \citenamefont {Antal}, \citenamefont {Szabó},\ and\ \citenamefont
  {Droz}}]{szabo_spatial_2000}%
  \BibitemOpen
  \bibfield  {author} {\bibinfo {author} {\bibfnamefont {G.}~\bibnamefont
  {Szabó}}, \bibinfo {author} {\bibfnamefont {T.}~\bibnamefont {Antal}},
  \bibinfo {author} {\bibfnamefont {P.}~\bibnamefont {Szabó}}, \ and\ \bibinfo
  {author} {\bibfnamefont {M.}~\bibnamefont {Droz}},\ }\href {\doibase
  10.1103/PhysRevE.62.1095} {\bibfield  {journal} {\bibinfo  {journal}
  {Physical Review E}\ }\textbf {\bibinfo {volume} {62}},\ \bibinfo {pages}
  {1095} (\bibinfo {year} {2000})}\BibitemShut {NoStop}%
\bibitem [{\citenamefont {Vukov}\ \emph {et~al.}(2006)\citenamefont {Vukov},
  \citenamefont {Szabó},\ and\ \citenamefont
  {Szolnoki}}]{vukov_cooperation_2006}%
  \BibitemOpen
  \bibfield  {author} {\bibinfo {author} {\bibfnamefont {J.}~\bibnamefont
  {Vukov}}, \bibinfo {author} {\bibfnamefont {G.}~\bibnamefont {Szabó}}, \
  and\ \bibinfo {author} {\bibfnamefont {A.}~\bibnamefont {Szolnoki}},\ }\href
  {\doibase 10.1103/PhysRevE.73.067103} {\bibfield  {journal} {\bibinfo
  {journal} {Physical Review E}\ }\textbf {\bibinfo {volume} {73}},\ \bibinfo
  {pages} {067103} (\bibinfo {year} {2006})}\BibitemShut {NoStop}%
\bibitem [{\citenamefont {Amaral}\ \emph {et~al.}(2016)\citenamefont {Amaral},
  \citenamefont {Wardil}, \citenamefont {Perc},\ and\ \citenamefont
  {da~Silva}}]{amaral_stochastic_2016}%
  \BibitemOpen
  \bibfield  {author} {\bibinfo {author} {\bibfnamefont {M.~A.}\ \bibnamefont
  {Amaral}}, \bibinfo {author} {\bibfnamefont {L.}~\bibnamefont {Wardil}},
  \bibinfo {author} {\bibfnamefont {M.}~\bibnamefont {Perc}}, \ and\ \bibinfo
  {author} {\bibfnamefont {J.~K.~L.}\ \bibnamefont {da~Silva}},\ }\href
  {\doibase 10.1103/PhysRevE.94.032317} {\bibfield  {journal} {\bibinfo
  {journal} {Physical Review E}\ }\textbf {\bibinfo {volume} {94}},\ \bibinfo
  {pages} {032317} (\bibinfo {year} {2016})}\BibitemShut {NoStop}%
\bibitem [{\citenamefont {Enquist}\ and\ \citenamefont
  {Leimar}(1993)}]{enquist_evolution_1993}%
  \BibitemOpen
  \bibfield  {author} {\bibinfo {author} {\bibfnamefont {M.}~\bibnamefont
  {Enquist}}\ and\ \bibinfo {author} {\bibfnamefont {O.}~\bibnamefont
  {Leimar}},\ }\href {\doibase 10.1006/anbe.1993.1089} {\bibfield  {journal}
  {\bibinfo  {journal} {Animal Behaviour}\ }\textbf {\bibinfo {volume} {45}},\
  \bibinfo {pages} {747} (\bibinfo {year} {1993})}\BibitemShut {NoStop}%
\bibitem [{\citenamefont {Ferrière}\ and\ \citenamefont
  {Michod}(1995)}]{ferriere_invading_1995}%
  \BibitemOpen
  \bibfield  {author} {\bibinfo {author} {\bibfnamefont {R.}~\bibnamefont
  {Ferrière}}\ and\ \bibinfo {author} {\bibfnamefont {R.~E.}\ \bibnamefont
  {Michod}},\ }\href {\doibase 10.1098/rspb.1995.0012} {\bibfield  {journal}
  {\bibinfo  {journal} {Proceedings of the Royal Society of London. Series B:
  Biological Sciences}\ }\textbf {\bibinfo {volume} {259}},\ \bibinfo {pages}
  {77} (\bibinfo {year} {1995})}\BibitemShut {NoStop}%
\bibitem [{\citenamefont {Aktipis}(2004)}]{aktipis_know_2004}%
  \BibitemOpen
  \bibfield  {author} {\bibinfo {author} {\bibfnamefont {C.~A.}\ \bibnamefont
  {Aktipis}},\ }\href {\doibase 10.1016/j.jtbi.2004.06.020} {\bibfield
  {journal} {\bibinfo  {journal} {Journal of Theoretical Biology}\ }\textbf
  {\bibinfo {volume} {231}},\ \bibinfo {pages} {249} (\bibinfo {year}
  {2004})}\BibitemShut {NoStop}%
\bibitem [{\citenamefont {Hamilton}\ and\ \citenamefont
  {Taborsky}(2005)}]{hamilton_contingent_2005}%
  \BibitemOpen
  \bibfield  {author} {\bibinfo {author} {\bibfnamefont {I.~M.}\ \bibnamefont
  {Hamilton}}\ and\ \bibinfo {author} {\bibfnamefont {M.}~\bibnamefont
  {Taborsky}},\ }\href {https://www.jstor.org/stable/30047816} {\bibfield
  {journal} {\bibinfo  {journal} {Proceedings: Biological Sciences}\ }\textbf
  {\bibinfo {volume} {272}},\ \bibinfo {pages} {2259} (\bibinfo {year}
  {2005})}\BibitemShut {NoStop}%
\bibitem [{\citenamefont {Le~Galliard}\ \emph {et~al.}(2005)\citenamefont
  {Le~Galliard}, \citenamefont {Ferrière},\ and\ \citenamefont
  {Dieckmann}}]{le_galliard_adaptive_2005}%
  \BibitemOpen
  \bibfield  {author} {\bibinfo {author} {\bibfnamefont {J.}~\bibnamefont
  {Le~Galliard}}, \bibinfo {author} {\bibfnamefont {R.}~\bibnamefont
  {Ferrière}}, \ and\ \bibinfo {author} {\bibfnamefont {U.}~\bibnamefont
  {Dieckmann}},\ }\href {\doibase 10.1086/427090} {\bibfield  {journal}
  {\bibinfo  {journal} {The American Naturalist}\ }\textbf {\bibinfo {volume}
  {165}},\ \bibinfo {pages} {206} (\bibinfo {year} {2005})}\BibitemShut
  {NoStop}%
\bibitem [{\citenamefont {Helbing}\ and\ \citenamefont
  {Yu}(2009)}]{helbing_outbreak_2009}%
  \BibitemOpen
  \bibfield  {author} {\bibinfo {author} {\bibfnamefont {D.}~\bibnamefont
  {Helbing}}\ and\ \bibinfo {author} {\bibfnamefont {W.}~\bibnamefont {Yu}},\
  }\href {\doibase 10.1073/pnas.0811503106} {\bibfield  {journal} {\bibinfo
  {journal} {Proceedings of the National Academy of Sciences}\ }\textbf
  {\bibinfo {volume} {106}},\ \bibinfo {pages} {3680} (\bibinfo {year}
  {2009})}\BibitemShut {NoStop}%
\bibitem [{\citenamefont {Meloni}\ \emph {et~al.}(2009)\citenamefont {Meloni},
  \citenamefont {Buscarino}, \citenamefont {Fortuna}, \citenamefont {Frasca},
  \citenamefont {Gomez-Gardenes}, \citenamefont {Latora},\ and\ \citenamefont
  {Moreno}}]{meloni_effects_2009}%
  \BibitemOpen
  \bibfield  {author} {\bibinfo {author} {\bibfnamefont {S.}~\bibnamefont
  {Meloni}}, \bibinfo {author} {\bibfnamefont {A.}~\bibnamefont {Buscarino}},
  \bibinfo {author} {\bibfnamefont {L.}~\bibnamefont {Fortuna}}, \bibinfo
  {author} {\bibfnamefont {M.}~\bibnamefont {Frasca}}, \bibinfo {author}
  {\bibfnamefont {J.}~\bibnamefont {Gomez-Gardenes}}, \bibinfo {author}
  {\bibfnamefont {V.}~\bibnamefont {Latora}}, \ and\ \bibinfo {author}
  {\bibfnamefont {Y.}~\bibnamefont {Moreno}},\ }\href {\doibase
  10.1103/PhysRevE.79.067101} {\bibfield  {journal} {\bibinfo  {journal}
  {Physical Review E}\ }\textbf {\bibinfo {volume} {79}},\ \bibinfo {pages}
  {067101} (\bibinfo {year} {2009})}\BibitemShut {NoStop}%
\bibitem [{\citenamefont {Smaldino}\ and\ \citenamefont
  {Schank}(2012)}]{smaldino_movement_2012}%
  \BibitemOpen
  \bibfield  {author} {\bibinfo {author} {\bibfnamefont {P.~E.}\ \bibnamefont
  {Smaldino}}\ and\ \bibinfo {author} {\bibfnamefont {J.~C.}\ \bibnamefont
  {Schank}},\ }\href {\doibase 10.1016/j.tpb.2012.03.004} {\bibfield  {journal}
  {\bibinfo  {journal} {Theoretical Population Biology}\ }\textbf {\bibinfo
  {volume} {82}},\ \bibinfo {pages} {48} (\bibinfo {year} {2012})}\BibitemShut
  {NoStop}%
\bibitem [{\citenamefont {Yang}\ and\ \citenamefont
  {Zhang}(2023)}]{yang_random_2023}%
  \BibitemOpen
  \bibfield  {author} {\bibinfo {author} {\bibfnamefont {Z.}~\bibnamefont
  {Yang}}\ and\ \bibinfo {author} {\bibfnamefont {L.}~\bibnamefont {Zhang}},\
  }\href {\doibase 10.1063/5.0139874} {\bibfield  {journal} {\bibinfo
  {journal} {Chaos: An Interdisciplinary Journal of Nonlinear Science}\
  }\textbf {\bibinfo {volume} {33}},\ \bibinfo {pages} {043126} (\bibinfo
  {year} {2023})}\BibitemShut {NoStop}%
\bibitem [{\citenamefont {Ben-Jacob}\ \emph {et~al.}(2000)\citenamefont
  {Ben-Jacob}, \citenamefont {Cohen},\ and\ \citenamefont
  {Levine}}]{ben-jacob_cooperative_2000}%
  \BibitemOpen
  \bibfield  {author} {\bibinfo {author} {\bibfnamefont {E.}~\bibnamefont
  {Ben-Jacob}}, \bibinfo {author} {\bibfnamefont {I.}~\bibnamefont {Cohen}}, \
  and\ \bibinfo {author} {\bibfnamefont {H.}~\bibnamefont {Levine}},\ }\href
  {\doibase 10.1080/000187300405228} {\bibfield  {journal} {\bibinfo  {journal}
  {Advances in Physics}\ }\textbf {\bibinfo {volume} {49}},\ \bibinfo {pages}
  {395} (\bibinfo {year} {2000})}\BibitemShut {NoStop}%
\bibitem [{\citenamefont {Bryant}\ and\ \citenamefont
  {Machta}(2023)}]{bryant_physical_2023}%
  \BibitemOpen
  \bibfield  {author} {\bibinfo {author} {\bibfnamefont {S.~J.}\ \bibnamefont
  {Bryant}}\ and\ \bibinfo {author} {\bibfnamefont {B.~B.}\ \bibnamefont
  {Machta}},\ }\href {\doibase 10.1103/PhysRevLett.131.068401} {\bibfield
  {journal} {\bibinfo  {journal} {Physical Review Letters}\ }\textbf {\bibinfo
  {volume} {131}},\ \bibinfo {pages} {068401} (\bibinfo {year}
  {2023})}\BibitemShut {NoStop}%
\bibitem [{\citenamefont {Hölldobler}\ and\ \citenamefont
  {Wilson}(1990)}]{holldobler_ants_1990}%
  \BibitemOpen
  \bibfield  {author} {\bibinfo {author} {\bibfnamefont {B.}~\bibnamefont
  {Hölldobler}}\ and\ \bibinfo {author} {\bibfnamefont {E.~O.}\ \bibnamefont
  {Wilson}},\ }\href@noop {} {\emph {\bibinfo {title} {The ants}}}\ (\bibinfo
  {publisher} {Springer},\ \bibinfo {address} {Berlin},\ \bibinfo {year}
  {1990})\BibitemShut {NoStop}%
\bibitem [{\citenamefont {Cattivelli}\ and\ \citenamefont
  {Sayed}(2011)}]{cattivelli_modeling_2011}%
  \BibitemOpen
  \bibfield  {author} {\bibinfo {author} {\bibfnamefont {F.~S.}\ \bibnamefont
  {Cattivelli}}\ and\ \bibinfo {author} {\bibfnamefont {A.~H.}\ \bibnamefont
  {Sayed}},\ }\href {\doibase 10.1109/TSP.2011.2107907} {\bibfield  {journal}
  {\bibinfo  {journal} {IEEE Transactions on Signal Processing}\ }\textbf
  {\bibinfo {volume} {59}},\ \bibinfo {pages} {2038} (\bibinfo {year}
  {2011})}\BibitemShut {NoStop}%
\bibitem [{\citenamefont {Turing}(1952)}]{turing_chemical_1952}%
  \BibitemOpen
  \bibfield  {author} {\bibinfo {author} {\bibfnamefont {A.~M.}\ \bibnamefont
  {Turing}},\ }\href {\doibase 10.1098/rstb.1952.0012} {\bibfield  {journal}
  {\bibinfo  {journal} {Philosophical Transactions of the Royal Society of
  London. Series B, Biological Sciences}\ }\textbf {\bibinfo {volume} {237}},\
  \bibinfo {pages} {37} (\bibinfo {year} {1952})}\BibitemShut {NoStop}%
\bibitem [{\citenamefont {Nagel}\ and\ \citenamefont
  {Paczuski}(1995)}]{nagel_emergent_1995}%
  \BibitemOpen
  \bibfield  {author} {\bibinfo {author} {\bibfnamefont {K.}~\bibnamefont
  {Nagel}}\ and\ \bibinfo {author} {\bibfnamefont {M.}~\bibnamefont
  {Paczuski}},\ }\href {\doibase 10.1103/PhysRevE.51.2909} {\bibfield
  {journal} {\bibinfo  {journal} {Physical Review E}\ }\textbf {\bibinfo
  {volume} {51}},\ \bibinfo {pages} {2909} (\bibinfo {year}
  {1995})}\BibitemShut {NoStop}%
\bibitem [{\citenamefont {von Schantz}\ and\ \citenamefont
  {Ehtamo}(2015)}]{von_schantz_spatial_2015}%
  \BibitemOpen
  \bibfield  {author} {\bibinfo {author} {\bibfnamefont {A.}~\bibnamefont {von
  Schantz}}\ and\ \bibinfo {author} {\bibfnamefont {H.}~\bibnamefont
  {Ehtamo}},\ }\href {\doibase 10.1103/PhysRevE.92.052805} {\bibfield
  {journal} {\bibinfo  {journal} {Physical Review E}\ }\textbf {\bibinfo
  {volume} {92}},\ \bibinfo {pages} {052805} (\bibinfo {year}
  {2015})}\BibitemShut {NoStop}%
\bibitem [{\citenamefont {Yanagisawa}(2016)}]{yanagisawa_coordination_2016}%
  \BibitemOpen
  \bibfield  {author} {\bibinfo {author} {\bibfnamefont {D.}~\bibnamefont
  {Yanagisawa}},\ }\href {\doibase 10.17815/CD.2016.8} {\bibfield  {journal}
  {\bibinfo  {journal} {Collective Dynamics}\ }\textbf {\bibinfo {volume}
  {1}},\ \bibinfo {pages} {1} (\bibinfo {year} {2016})}\BibitemShut {NoStop}%
\bibitem [{\citenamefont {Ferriere}\ and\ \citenamefont
  {Michod}(1996)}]{ferriere_evolution_1996}%
  \BibitemOpen
  \bibfield  {author} {\bibinfo {author} {\bibfnamefont {R.}~\bibnamefont
  {Ferriere}}\ and\ \bibinfo {author} {\bibfnamefont {R.~E.}\ \bibnamefont
  {Michod}},\ }\href {\doibase 10.1086/285875} {\bibfield  {journal} {\bibinfo
  {journal} {The American Naturalist}\ }\textbf {\bibinfo {volume} {147}},\
  \bibinfo {pages} {692} (\bibinfo {year} {1996})}\BibitemShut {NoStop}%
\bibitem [{\citenamefont {Vainstein}\ \emph {et~al.}(2007)\citenamefont
  {Vainstein}, \citenamefont {Silva},\ and\ \citenamefont
  {Arenzon}}]{vainstein_does_2007}%
  \BibitemOpen
  \bibfield  {author} {\bibinfo {author} {\bibfnamefont {M.~H.}\ \bibnamefont
  {Vainstein}}, \bibinfo {author} {\bibfnamefont {A.~T.~C.}\ \bibnamefont
  {Silva}}, \ and\ \bibinfo {author} {\bibfnamefont {J.~J.}\ \bibnamefont
  {Arenzon}},\ }\href {\doibase 10.1016/j.jtbi.2006.09.012} {\bibfield
  {journal} {\bibinfo  {journal} {Journal of Theoretical Biology}\ }\textbf
  {\bibinfo {volume} {244}},\ \bibinfo {pages} {722} (\bibinfo {year}
  {2007})}\BibitemShut {NoStop}%
\bibitem [{\citenamefont {Sicardi}\ \emph {et~al.}(2009)\citenamefont
  {Sicardi}, \citenamefont {Fort}, \citenamefont {Vainstein},\ and\
  \citenamefont {Arenzon}}]{sicardi_random_2009}%
  \BibitemOpen
  \bibfield  {author} {\bibinfo {author} {\bibfnamefont {E.~A.}\ \bibnamefont
  {Sicardi}}, \bibinfo {author} {\bibfnamefont {H.}~\bibnamefont {Fort}},
  \bibinfo {author} {\bibfnamefont {M.~H.}\ \bibnamefont {Vainstein}}, \ and\
  \bibinfo {author} {\bibfnamefont {J.~J.}\ \bibnamefont {Arenzon}},\ }\href
  {\doibase 10.1016/j.jtbi.2008.09.022} {\bibfield  {journal} {\bibinfo
  {journal} {Journal of Theoretical Biology}\ }\textbf {\bibinfo {volume}
  {256}},\ \bibinfo {pages} {240} (\bibinfo {year} {2009})}\BibitemShut
  {NoStop}%
\bibitem [{\citenamefont {Droz}\ \emph {et~al.}(2009)\citenamefont {Droz},
  \citenamefont {Szwabiński},\ and\ \citenamefont
  {Szabó}}]{droz_motion_2009}%
  \BibitemOpen
  \bibfield  {author} {\bibinfo {author} {\bibfnamefont {M.}~\bibnamefont
  {Droz}}, \bibinfo {author} {\bibfnamefont {J.}~\bibnamefont {Szwabiński}}, \
  and\ \bibinfo {author} {\bibfnamefont {G.}~\bibnamefont {Szabó}},\ }\href
  {\doibase 10.1140/epjb/e2009-00160-1} {\bibfield  {journal} {\bibinfo
  {journal} {The European Physical Journal B}\ }\textbf {\bibinfo {volume}
  {71}},\ \bibinfo {pages} {579} (\bibinfo {year} {2009})}\BibitemShut
  {NoStop}%
\bibitem [{\citenamefont {Cheng}\ \emph {et~al.}(2010)\citenamefont {Cheng},
  \citenamefont {Li}, \citenamefont {Dai}, \citenamefont {Zhu},\ and\
  \citenamefont {Yang}}]{cheng_motion_2010}%
  \BibitemOpen
  \bibfield  {author} {\bibinfo {author} {\bibfnamefont {H.}~\bibnamefont
  {Cheng}}, \bibinfo {author} {\bibfnamefont {H.}~\bibnamefont {Li}}, \bibinfo
  {author} {\bibfnamefont {Q.}~\bibnamefont {Dai}}, \bibinfo {author}
  {\bibfnamefont {Y.}~\bibnamefont {Zhu}}, \ and\ \bibinfo {author}
  {\bibfnamefont {J.}~\bibnamefont {Yang}},\ }\href {\doibase
  10.1088/1367-2630/12/12/123014} {\bibfield  {journal} {\bibinfo  {journal}
  {New Journal of Physics}\ }\textbf {\bibinfo {volume} {12}},\ \bibinfo
  {pages} {123014} (\bibinfo {year} {2010})}\BibitemShut {NoStop}%
\bibitem [{\citenamefont {Suzuki}\ and\ \citenamefont
  {Kimura}(2011)}]{suzuki_oscillatory_2011}%
  \BibitemOpen
  \bibfield  {author} {\bibinfo {author} {\bibfnamefont {S.}~\bibnamefont
  {Suzuki}}\ and\ \bibinfo {author} {\bibfnamefont {H.}~\bibnamefont
  {Kimura}},\ }\href {\doibase 10.1016/j.jtbi.2011.07.019} {\bibfield
  {journal} {\bibinfo  {journal} {Journal of Theoretical Biology}\ }\textbf
  {\bibinfo {volume} {287}},\ \bibinfo {pages} {42} (\bibinfo {year}
  {2011})}\BibitemShut {NoStop}%
\bibitem [{\citenamefont {Yang}\ and\ \citenamefont
  {Wang}(2011)}]{yang_universal_2011}%
  \BibitemOpen
  \bibfield  {author} {\bibinfo {author} {\bibfnamefont {H.}~\bibnamefont
  {Yang}}\ and\ \bibinfo {author} {\bibfnamefont {B.}~\bibnamefont {Wang}},\
  }\href {\doibase 10.1007/s11434-011-4768-5} {\bibfield  {journal} {\bibinfo
  {journal} {Chinese Science Bulletin}\ }\textbf {\bibinfo {volume} {56}},\
  \bibinfo {pages} {3693} (\bibinfo {year} {2011})}\BibitemShut {NoStop}%
\bibitem [{\citenamefont {Gelimson}\ \emph {et~al.}(2013)\citenamefont
  {Gelimson}, \citenamefont {Cremer},\ and\ \citenamefont
  {Frey}}]{gelimson_mobility_2013}%
  \BibitemOpen
  \bibfield  {author} {\bibinfo {author} {\bibfnamefont {A.}~\bibnamefont
  {Gelimson}}, \bibinfo {author} {\bibfnamefont {J.}~\bibnamefont {Cremer}}, \
  and\ \bibinfo {author} {\bibfnamefont {E.}~\bibnamefont {Frey}},\ }\href
  {\doibase 10.1103/PhysRevE.87.042711} {\bibfield  {journal} {\bibinfo
  {journal} {Physical Review E}\ }\textbf {\bibinfo {volume} {87}},\ \bibinfo
  {pages} {042711} (\bibinfo {year} {2013})}\BibitemShut {NoStop}%
\bibitem [{\citenamefont {Vainstein}\ \emph {et~al.}(2014)\citenamefont
  {Vainstein}, \citenamefont {Brito},\ and\ \citenamefont
  {Arenzon}}]{vainstein_percolation_2014}%
  \BibitemOpen
  \bibfield  {author} {\bibinfo {author} {\bibfnamefont {M.~H.}\ \bibnamefont
  {Vainstein}}, \bibinfo {author} {\bibfnamefont {C.}~\bibnamefont {Brito}}, \
  and\ \bibinfo {author} {\bibfnamefont {J.~J.}\ \bibnamefont {Arenzon}},\
  }\href {\doibase 10.1103/PhysRevE.90.022132} {\bibfield  {journal} {\bibinfo
  {journal} {Physical Review E}\ }\textbf {\bibinfo {volume} {90}},\ \bibinfo
  {pages} {022132} (\bibinfo {year} {2014})}\BibitemShut {NoStop}%
\bibitem [{\citenamefont {Fu}\ \emph {et~al.}(2010)\citenamefont {Fu},
  \citenamefont {Nowak},\ and\ \citenamefont {Hauert}}]{fu_invasion_2010}%
  \BibitemOpen
  \bibfield  {author} {\bibinfo {author} {\bibfnamefont {F.}~\bibnamefont
  {Fu}}, \bibinfo {author} {\bibfnamefont {M.~A.}\ \bibnamefont {Nowak}}, \
  and\ \bibinfo {author} {\bibfnamefont {C.}~\bibnamefont {Hauert}},\ }\href
  {\doibase 10.1016/j.jtbi.2010.06.042} {\bibfield  {journal} {\bibinfo
  {journal} {Journal of Theoretical Biology}\ }\textbf {\bibinfo {volume}
  {266}},\ \bibinfo {pages} {358} (\bibinfo {year} {2010})}\BibitemShut
  {NoStop}%
\bibitem [{\citenamefont {Lindgren}\ and\ \citenamefont
  {Nordahl}(1994)}]{lindgren_evolutionary_1994}%
  \BibitemOpen
  \bibfield  {author} {\bibinfo {author} {\bibfnamefont {K.}~\bibnamefont
  {Lindgren}}\ and\ \bibinfo {author} {\bibfnamefont {M.~G.}\ \bibnamefont
  {Nordahl}},\ }\href {\doibase 10.1016/0167-2789(94)90289-5} {\bibfield
  {journal} {\bibinfo  {journal} {Physica D: Nonlinear Phenomena}\ }\textbf
  {\bibinfo {volume} {75}},\ \bibinfo {pages} {292} (\bibinfo {year}
  {1994})}\BibitemShut {NoStop}%
\bibitem [{\citenamefont {Nowak}\ and\ \citenamefont
  {Sigmund}(1999)}]{nowak_games_1999}%
  \BibitemOpen
  \bibfield  {author} {\bibinfo {author} {\bibfnamefont {M.~A.}\ \bibnamefont
  {Nowak}}\ and\ \bibinfo {author} {\bibfnamefont {K.}~\bibnamefont
  {Sigmund}},\ }in\ \href@noop {} {\emph {\bibinfo {booktitle} {The {Geometry}
  of {Ecological} {Interactions}: {Simplifying} {Spatial} {Complexity}}}}\
  (\bibinfo {year} {1999})\BibitemShut {NoStop}%
\bibitem [{\citenamefont {Fisher}(1937)}]{fisher_wave_1937}%
  \BibitemOpen
  \bibfield  {author} {\bibinfo {author} {\bibfnamefont {R.~A.}\ \bibnamefont
  {Fisher}},\ }\href {\doibase 10.1111/j.1469-1809.1937.tb02153.x} {\bibfield
  {journal} {\bibinfo  {journal} {Annals of Eugenics}\ }\textbf {\bibinfo
  {volume} {7}},\ \bibinfo {pages} {355} (\bibinfo {year} {1937})}\BibitemShut
  {NoStop}%
\bibitem [{\citenamefont {Flores}\ \emph {et~al.}(2022)\citenamefont {Flores},
  \citenamefont {Amaral}, \citenamefont {Vainstein},\ and\ \citenamefont
  {Fernandes}}]{flores_cooperation_2022}%
  \BibitemOpen
  \bibfield  {author} {\bibinfo {author} {\bibfnamefont {L.~S.}\ \bibnamefont
  {Flores}}, \bibinfo {author} {\bibfnamefont {M.~A.}\ \bibnamefont {Amaral}},
  \bibinfo {author} {\bibfnamefont {M.~H.}\ \bibnamefont {Vainstein}}, \ and\
  \bibinfo {author} {\bibfnamefont {H.~C.~M.}\ \bibnamefont {Fernandes}},\
  }\href {\doibase 10.1016/j.chaos.2022.112744} {\enquote {\bibinfo {title}
  {Cooperation in regular lattices},}\ } (\bibinfo {year} {2022})\BibitemShut
  {NoStop}%
\bibitem [{\citenamefont {Nowak}\ \emph {et~al.}(1994)\citenamefont {Nowak},
  \citenamefont {Bonhoeffer},\ and\ \citenamefont {May}}]{nowak_spatial_1994}%
  \BibitemOpen
  \bibfield  {author} {\bibinfo {author} {\bibfnamefont {M.~A.}\ \bibnamefont
  {Nowak}}, \bibinfo {author} {\bibfnamefont {S.}~\bibnamefont {Bonhoeffer}}, \
  and\ \bibinfo {author} {\bibfnamefont {R.~M.}\ \bibnamefont {May}},\ }\href
  {\doibase 10.1073/pnas.91.11.4877} {\bibfield  {journal} {\bibinfo  {journal}
  {Proceedings of the National Academy of Sciences}\ }\textbf {\bibinfo
  {volume} {91}},\ \bibinfo {pages} {4877} (\bibinfo {year}
  {1994})}\BibitemShut {NoStop}%
\bibitem [{\citenamefont {Wang}\ \emph {et~al.}(2012)\citenamefont {Wang},
  \citenamefont {Xia}, \citenamefont {Wang}, \citenamefont {Ding},\ and\
  \citenamefont {Sun}}]{wang_spatial_2012}%
  \BibitemOpen
  \bibfield  {author} {\bibinfo {author} {\bibfnamefont {J.}~\bibnamefont
  {Wang}}, \bibinfo {author} {\bibfnamefont {C.}~\bibnamefont {Xia}}, \bibinfo
  {author} {\bibfnamefont {Y.}~\bibnamefont {Wang}}, \bibinfo {author}
  {\bibfnamefont {S.}~\bibnamefont {Ding}}, \ and\ \bibinfo {author}
  {\bibfnamefont {J.}~\bibnamefont {Sun}},\ }\href {\doibase
  10.1007/s11434-011-4890-4} {\bibfield  {journal} {\bibinfo  {journal}
  {Chinese Science Bulletin}\ }\textbf {\bibinfo {volume} {57}},\ \bibinfo
  {pages} {724} (\bibinfo {year} {2012})}\BibitemShut {NoStop}%
\bibitem [{\citenamefont {Szabo}\ and\ \citenamefont
  {Fath}(2007)}]{szabo_evolutionary_2007}%
  \BibitemOpen
  \bibfield  {author} {\bibinfo {author} {\bibfnamefont {G.}~\bibnamefont
  {Szabo}}\ and\ \bibinfo {author} {\bibfnamefont {G.}~\bibnamefont {Fath}},\
  }\href {\doibase 10.1016/j.physrep.2007.04.004} {\bibfield  {journal}
  {\bibinfo  {journal} {Physics Reports}\ }\textbf {\bibinfo {volume} {446}},\
  \bibinfo {pages} {97} (\bibinfo {year} {2007})}\BibitemShut {NoStop}%
\bibitem [{\citenamefont {Szolnoki}\ \emph {et~al.}(2014)\citenamefont
  {Szolnoki}, \citenamefont {Mobilia}, \citenamefont {Jiang}, \citenamefont
  {Szczesny}, \citenamefont {Rucklidge},\ and\ \citenamefont
  {Perc}}]{szolnoki_cyclic_2014}%
  \BibitemOpen
  \bibfield  {author} {\bibinfo {author} {\bibfnamefont {A.}~\bibnamefont
  {Szolnoki}}, \bibinfo {author} {\bibfnamefont {M.}~\bibnamefont {Mobilia}},
  \bibinfo {author} {\bibfnamefont {L.-L.}\ \bibnamefont {Jiang}}, \bibinfo
  {author} {\bibfnamefont {B.}~\bibnamefont {Szczesny}}, \bibinfo {author}
  {\bibfnamefont {A.~M.}\ \bibnamefont {Rucklidge}}, \ and\ \bibinfo {author}
  {\bibfnamefont {M.}~\bibnamefont {Perc}},\ }\href {\doibase
  10.1098/rsif.2014.0735} {\bibfield  {journal} {\bibinfo  {journal} {Journal
  of The Royal Society Interface}\ }\textbf {\bibinfo {volume} {11}},\ \bibinfo
  {pages} {20140735} (\bibinfo {year} {2014})}\BibitemShut {NoStop}%
\bibitem [{\citenamefont {Song}\ \emph {et~al.}(2015)\citenamefont {Song},
  \citenamefont {Gokhale}, \citenamefont {Papkou}, \citenamefont
  {Schulenburg},\ and\ \citenamefont {Traulsen}}]{song_host-parasite_2015}%
  \BibitemOpen
  \bibfield  {author} {\bibinfo {author} {\bibfnamefont {Y.}~\bibnamefont
  {Song}}, \bibinfo {author} {\bibfnamefont {C.~S.}\ \bibnamefont {Gokhale}},
  \bibinfo {author} {\bibfnamefont {A.}~\bibnamefont {Papkou}}, \bibinfo
  {author} {\bibfnamefont {H.}~\bibnamefont {Schulenburg}}, \ and\ \bibinfo
  {author} {\bibfnamefont {A.}~\bibnamefont {Traulsen}},\ }\href {\doibase
  10.1186/s12862-015-0462-6} {\bibfield  {journal} {\bibinfo  {journal} {BMC
  Evolutionary Biology}\ }\textbf {\bibinfo {volume} {15}} (\bibinfo {year}
  {2015}),\ 10.1186/s12862-015-0462-6}\BibitemShut {NoStop}%
\bibitem [{\citenamefont {Kondepudi}\ and\ \citenamefont
  {Prigogine}(2015)}]{kondepudi_modern_2015}%
  \BibitemOpen
  \bibfield  {author} {\bibinfo {author} {\bibfnamefont {D.}~\bibnamefont
  {Kondepudi}}\ and\ \bibinfo {author} {\bibfnamefont {I.}~\bibnamefont
  {Prigogine}},\ }\href@noop {} {\emph {\bibinfo {title} {Modern
  thermodynamics: from heat engines to dissipative structures}}}\ (\bibinfo
  {year} {2015})\BibitemShut {NoStop}%
\bibitem [{\citenamefont {Vallis}(2017)}]{vallis_atmospheric_2017}%
  \BibitemOpen
  \bibfield  {author} {\bibinfo {author} {\bibfnamefont {G.~K.}\ \bibnamefont
  {Vallis}},\ }\href
  {https://www.cambridge.org/core/books/atmospheric-and-oceanic-fluid-dynamics/41379BDDC4257CBE11143C466F6428A4}
  {\emph {\bibinfo {title} {Atmospheric and {Oceanic} {Fluid} {Dynamics}:
  {Fundamentals} and {Large}-{Scale} {Circulation}}}},\ \bibinfo {edition}
  {2nd}\ ed.\ (\bibinfo  {publisher} {Cambridge University Press},\ \bibinfo
  {address} {Cambridge},\ \bibinfo {year} {2017})\BibitemShut {NoStop}%
\end{thebibliography}
%

\end{document}